\pdfoutput=1
\documentclass[a4paper,12pt]{article}
\usepackage{jheppub}

\usepackage{amsmath,amsfonts,amssymb,mathrsfs}
\usepackage{natbib}
\usepackage{enumitem}
\usepackage[latin1]{inputenc}
\usepackage{graphicx}
\usepackage[font=small,labelfont=bf]{caption}
\usepackage{subcaption}
\usepackage{tikz}
\usetikzlibrary{calc,quotes,positioning,shapes.misc,fit}

\usepackage[colorlinks=true]{hyperref} 
\hypersetup{
    bookmarks=true,         
    unicode=false,          
    pdftoolbar=true,        
    pdfmenubar=true,        
    pdffitwindow=false,     
    pdfstartview={FitH},    
    pdftitle={p-adic CFT is a Holographic Tensor Network},    
    pdfauthor={Ling-Yan Hung, Wei Li, and Charles M. Melby-Thompson},     
    pdfnewwindow=true,      
    colorlinks=true,        
    linkcolor=blue,         
    citecolor=red,          
    filecolor=magenta,      
    urlcolor=cyan           
} 

\def\pTN{}

\newcommand*\pgfdeclareanchoralias[3]{%
    \expandafter\def\csname pgf@anchor@#1@#3\expandafter\endcsname
            \expandafter{\csname pgf@anchor@#1@#2\endcsname}}

\def\ketsize{0.25}
\def\ketinput{\ketsize}
\def\ketshapemult{0.7}

\def\tensorsize{1.3em}
\def\propagatorsize{1.5em}

\tikzset{
  vertex/.style =     {text centered, inner sep=0pt, text width=\tensorsize},
  tensor/.style =     {vertex, circle, draw=black, fill=white, minimum size=\tensorsize},
  propagator/.style = {vertex, rectangle, draw=black, fill=white, 
                       minimum width=\propagatorsize, minimum height=\propagatorsize}
}

\pgfdeclareanchoralias{rounded rectangle}{west}{in}
\pgfdeclareanchoralias{rounded rectangle}{east}{out}

\tikzset{
  bra/.style = { rounded rectangle, 
                 rounded rectangle east arc=none,
                 draw           = black,
                 fill           = white,
                 inner xsep     = 2.5pt,
                 inner ysep     = 2pt,
                 minimum width  = 2em,
                 minimum height = 1.2em,
                 anchor         = east            },
  ket/.style = { rounded rectangle, 
                 rounded rectangle west arc=none,
                 draw           = black,
                 fill           = white,
                 inner xsep     = 2.5pt,
                 inner ysep     = 2pt,
                 minimum width  = 2em,
                 minimum height = 1.2em,
                 anchor         = west            }
}

\tikzset{
  ket/.pic ={
    \coordinate (-in) at (-\ketinput,0);
    \draw[pic actions] (-\ketsize,-\ketsize) -- (-\ketsize,\ketsize) 
    	-- (\ketshapemult*\ketsize,\ketsize) -- (1.3*\ketsize,0) 
		-- (\ketshapemult*\ketsize,-\ketsize) --cycle;
	\draw[pic actions] (-\ketinput,0) -- (-\ketsize,0);
    \node{ \tikzpictext};
  },
  bra/.pic ={
    \coordinate (-out) at (\ketinput,0);
    \draw[pic actions] (\ketsize,-\ketsize) -- (\ketsize,\ketsize) 
    	-- (-\ketshapemult*\ketsize,\ketsize) -- (-1.3*\ketsize,0) 
		-- (-\ketshapemult*\ketsize,-\ketsize) --cycle;
	\draw[pic actions] (\ketinput,0) -- (\ketsize,0);
    \node{ \tikzpictext};
  }
}

\newcommand\figref[1]	{Fig.~\ref{#1}}
\newcommand\secref[1]	{Sec.~\ref{#1}}

\newcommand{\be}	{\begin{equation}}
\newcommand{\ee}	{\end{equation}}

\newcommand{\p}	{\partial}

\renewcommand{\H}	{\mathbb{H}}
\newcommand{\F}	{\mathbb{F}}
\newcommand{\Q}	{\mathbb{Q}}
\newcommand{\K}	{\mathbb{K}}

\newcommand{\R}	{\mathbb{R}}
\newcommand{\C}	{\mathbb{C}}
\newcommand{\Z}	{\mathbb{Z}}
\renewcommand{\P}	{\mathbb{P}}

\newcommand{\cA}{\mathcal{A}}
\newcommand{\cC}{\mathcal{C}}

\newcommand{\cG}{\mathcal{G}}
\newcommand{\cH}{\mathcal{H}}
\newcommand{\cI}{\mathcal{I}}

\newcommand{\cO}{\mathcal{O}}
\newcommand{\cP}{\mathcal{P}}
\newcommand{\cS}{\mathcal{S}}
\newcommand{\cT}{\mathcal{T}}

\newcommand{\cZ}{\mathcal{Z}}

\newcommand{\bra}[1]	{\langle{#1}\vert}
\newcommand{\ket}[1]	{\vert{#1}\rangle}
\newcommand{\braket}[2]	{\langle{#1}\vert{#2}\rangle}
\newcommand{\corr}[1]	{\left\langle{#1}\right\rangle}

\newcommand{\End}	{\mathrm{End}}
\newcommand{\id}	{\mathrm{id}}

\newcommand\Group[1]	{\mathrm{#1}}
\newcommand{\SO}	{\Group{SO}}

\newcommand{\SL}	{\Group{SL}}

\newcommand{\PSL}	{\Group{PSL}}
\newcommand{\PGL}	{\Group{PGL}}
\newcommand{\PSU}	{\Group{PSU}}

\newcommand{\AdS}	{\mathrm{AdS}}
\newcommand{\CFT}	{\mathrm{CFT}}

\newcommand\q{\mathsf{q}}

\begin{document}

\title{\bf $p$-adic CFT is a holographic tensor network}

\author{Ling-Yan Hung$^{a,b,c}$,  Wei Li$^{d}$ and Charles M. Melby-Thompson$^{e}$} 

\affiliation{$^a$ State Key Laboratory of Surface Physics and Department of Physics,\\ 
\hspace*{0.3cm}Fudan University, 220 Handan Road, 200433 Shanghai, P.\ R.\ China}

\affiliation{$^b$ Department of Physics and Center for Field Theory and Particle Physics,\\
\hspace*{0.3cm}Fudan University, Handan Road, 200433 Shanghai, P.\ R.\ China}
        
\affiliation{$c$ Institute for Nanoelectronic devices and Quantum computing, Fudan University, 200433 Shanghai, China}

\affiliation{$^d$ Institute of Theoretical Physics,\\
\hspace*{0.3cm}Chinese Academy of Sciences, 100190 Beijing, P.\ R.\ China}

\affiliation{$^e$ Institut f\"{u}r Theoretische Physik und Astrophysik,\\
\hspace*{0.3cm}Julius-Maximilians-Universit\"{a}t W\"{u}rzburg, Am Hubland, 97074 W\"{u}rzburg, Germany}

\emailAdd{elektron.janethung@gmail.com, 
weili@itp.ac.cn, charlesmelby@gmail.com}
\abstract{The $p$-adic AdS/CFT correspondence relates a CFT living on the $p$-adic numbers to a system living on the Bruhat-Tits tree. 
Modifying our earlier proposal \cite{Bhattacharyya:2017aly} for a tensor network realization of $p$-adic AdS/CFT, we prove that the path integral of a $p$-adic CFT is equivalent to a tensor network on the Bruhat-Tits tree, in the sense that the tensor network reproduces all correlation functions of the $p$-adic CFT. 
Our rules give an explicit tensor network for any $p$-adic CFT (as axiomatized by Melzer), and can be applied not only to the $p$-adic plane, but also to compute any correlation functions on higher genus $p$-adic curves.
Finally, we apply them to define and study RG flows in $p$-adic CFTs, establishing in particular that any IR fixed point is itself a $p$-adic CFT. 
}

\maketitle
\flushbottom
\newpage

\section{Introduction}

The field of rational numbers $\mathbb{Q}$ can be completed with respect to two different types of norms \cite{Ostrowski}: 
(1) the Euclidean or \emph{Archimedean} norm, giving the real field $\mathbb{R}$; and 
(2) the $p$-adic or \emph{non-Archimedean} norm, for any prime $p$, yielding the $p$-adic number field $\mathbb{Q}_p$. 
Much of the importance of $p$-adic numbers in mathematics is derived from the local-global principle, which allows some problems in number theory to be solved over $\Q$ by simultaneously solving them over $\R$ and all the $\Q_p$. 
However, their unique analytic properties have also inspired the development of various physics-like models built around the $p$-adics, including $p$-adic versions of strings \cite{Freund:1987kt,Freund:1987ck,Brekke:1988dg}, as well as applications to quantum mechanics, stochastic systems, biology, and beyond (for some reviews see \cite{Rammal:1986zz,Brekke:1993gf,Dragovich:2017kge}). 

Recent work \cite{Gubser:2016guj,Heydeman:2016ldy} extended the use of $p$-adics in physics to a new arena: the AdS/CFT correspondence.
In the $p$-adic version, the boundary CFT lives not on $\R^n$ or $\C$, but on $\mathbb{Q}_p$%
\footnote{As in \cite{Gubser:2016guj,Heydeman:2016ldy}, in the bulk of the paper we will allow $\Q_p$ to be replaced by any unramified finite field extension.}
and has conformal symmetry group $\PGL(2,\Q_p)$ rather than $\SO(n+1,1)$ or $\PSL(2,\C)$. 
The role of bulk geometry is played by the Bruhat-Tits tree --- a $(p+1)$-valent tree with symmetry group $\PGL(2,\mathbb{Q}_p)$ whose asymptotic boundary is $\mathbb{Q}_p$ \cite{BruhatTits}. 
Unlike the standard AdS/CFT correspondence, the $p$-adic bulk geometry is discrete, and the theory living on it is a statistical lattice theory with scalar fields living at each vertex \cite{Gubser:2016guj,Heydeman:2016ldy,Bhattacharyya:2017aly}. 
A natural generalization of the holographic dictionary to this statistical system yields boundary correlation functions satisfying the behavior expected of a CFT over the $p$-adic numbers.

The idea of describing a physical model in terms of a discrete geometry capturing some of its symmetries is reminiscent of tensor networks. 
Tensor networks find their primary uses as an efficient way to represent and compute two types of objects: low-lying quantum states in local quantum systems, as in MERA \cite{Vidal:2007hda}; and the infrared behavior of statistical transition matrices, as with tensor network renormalization (TNR) (see e.g.\ \cite{Levin:2006jai,2015PhRvL.115r0405E,2015arXiv151204938Y,Bal:2017mht}). 
The notion that tensor networks may be related to the way in which holographic systems encode information was proposed in \cite{Swingle:2009bg}, leading to the study of holography-like tensor networks placed either directly on a discretized spatial slice of the bulk geometry~\cite{HaPPY}, or on kinematic space~\cite{Czech:2015kbp}.

\medskip

The $p$-adic AdS/CFT correspondence, like its counterpart over the real numbers, is a conjecture.
However, it was suggested in \cite{Bhattacharyya:2017aly} that it may be possible to prove $p$-adic AdS/CFT by realizing it as a tensor network from which both bulk and boundary quantities can be extracted.
In particular, both the bulk reconstruction formula and the computation of boundary correlators in terms of Witten diagrams naturally emerge from the tensor network formulation. 

However, the picture presented in \cite{Bhattacharyya:2017aly} was not entirely satisfactory. 
There, the tensor network comprises a wavefunction $\ket{\Psi}$ living on $\Q_p$, and observables are computed by taking expectation values $\bra{\Psi}\cO_1(x_1)\cdots\cO_k(x_k)\ket{\Psi}$ of operators acting on its legs. 
One then seeks to impose the axioms of $p$-adic CFT on the resulting correlation functions. 
Unfortunately, the wavefunction constraints are quadratic and contain a large number of constraints, making it difficult to solve them, if they can be solved at all.%
\footnote{Tensor networks were also interpreted in \cite{Heydeman:2018qty} as the wavefunction of a $p$-adic CFT, and used to study entanglement entropy.
The behavior of this entropy remains puzzling, and it is not clear whether the tree structure can recover the expected entanglement entropy of the dual theory.}

We take a different approach in this work, and use tensor networks to describe instead the \emph{Euclidean path integral} of $p$-adic CFT.%
\footnote{It was already noted in \cite{Bhattacharyya:2017aly} that the Witten diagrams naturally contain bulk-boundary propagators that treat the Bruhat-Tits tree as space-time, rather than as a time slice.}
In this construction, the constraints on the tensor network arising from the OPE are linear, not quadratic, and take such a simple form that they can be solved explicitly. 
We can thereby show that any $p$-adic CFT is equivalent to a tensor network on the Bruhat-Tits tree, in the sense that it reproduces all local correlation functions of this $p$-adic CFT. 
Applying our prescription to the graphs associated to higher genus curves further allows the computation of correlation functions of the $p$-adic CFT on higher genus $p$-adic curves. 
Finally, we apply our prescription to define and study renormalization flows in $p$-adic CFTs, showing in particular that any fixed point of the RG flow itself satisfies the axioms of $p$-adic CFT.

It is interesting to note that the action of our tensor networks on a basis of primaries coincides with expectation values of a Wilson line network covering the Bruhat-Tits tree \cite{Hung:2018mcn}, offering an alternative interpretation of the construction. 

\smallskip
Our paper is organized as follows.
We begin in \secref{sec:Qp and pCFT} with a brief review of the $p$-adic numbers and their application to $p$-adic CFT as formalized in~\cite{Melzer}, and to $p$-adic AdS/CFT as originally described in \cite{Gubser:2016guj,Heydeman:2016ldy}. 
\secref{sec:pCFT=TN} establishes the main claim of this paper --- that $p$-adic CFT is equivalent to a tensor network --- and shows how to construct the holographic dictionary in detail. 
\secref{sec:higher genus} uses the tensor network/$p$-adic CFT equivalence to formulate and study $p$-adic CFT on curves of genus $g>0$, as described either by Schottky uniformization or by cutting and sewing, which yields a simple prescription for the computation of partition and general correlation functions.
Finally, section~\ref{sec:rg-flow} applies our tensor network construction to the study of RG flows in $p$-adic CFT, and examines the properties of the IR fixed point in a particular example.

\section{A review of \texorpdfstring{$p$}{p}-adic CFT and the \texorpdfstring{$p$}{p}-adic AdS/CFT correspondence}
\label{sec:Qp and pCFT} 
To make our discussion self-contained, we begin with the few basic facts about $p$-adic numbers that are necessary to what follows.
We then review the axioms of $p$-adic CFT \cite{Melzer} used in this paper, and the $p$-adic AdS/CFT correspondence as introduced in \cite{Gubser:2016guj,Heydeman:2016ldy}.

\subsection{\texorpdfstring{$p$}{p}-adic numbers and projective space}
\label{sec:Qp}
The $p$-adic numbers $\Q_p$ are the completion of the field $\Q$ of rational numbers with respect to the $p$-adic norm $|\cdot|_p$.
This is defined as follows: writing any rational number as $r=p^k a/b$ where $a$ and $b$ are integers relatively prime to $p$, set
\be
|r|_p = p^{-k} \,,
\qquad
|0|_p = 0 \,.
\ee
The $p$-adic norm makes $\Q_p$ a metric space, with metric $(x,y)_p=|x-y|_p$.
Like the standard metric $(x,y)_\infty=|x-y|$ on $\R$, $(\cdot,\cdot)_p$ satisfies the triangle inequality $(x,z) \le (x,y) + (y,z)$ for any $y$.
Unlike the real numbers, however, $(\cdot,\cdot)_p$ further satisfies the \textit{strong triangle} or \textit{ultrametric inequality}
\be
(x,z) \le \max\big[ (x,y),(y,z) \big].
\ee

A $p$-adic number $x$ with $p$-adic norm $p^{-k}$ can be expanded in powers of $p$,
\be
x = \sum_{n\ge k}x_n p^{n}
\ee
where each 
$x_n \in \F_p = \{0,\ldots,p-1\}$ and $x_k\ne 0$. 
Although such a series diverges in $\R$ when infinitely many $x_n\ne 0$, it always converges in $\Q_p$.

Given an unramified finite field extension $\K$ of $\Q_p$, the valuation can be extended to a valuation $|\cdot|_\K$ on $\K$.
With respect to multiplication by elements of $\Q_p$, such a $\K$ forms a finite-dimensional vector space isomorphic to $(\Q_p)^n$ where $n$ is the degree of the extension, making it natural to think of $n$ as the \emph{dimension} of $\K$. 
Any number in $\K$ also has an expansion in powers of $p$, where the coefficients are now valued in $\F_q$, the finite field of $q=p^n$ elements.

\newcommand\proj[2]{[{#1}\!:\!{#2}]}

A $p$-adic CFT lives most naturally on $p$-adic projective space $\P^1(\K)$ (or on curves locally isomorphic to it), where $\K$ is some $p$-adic field. 
We will often refer to $\P^1(\K)$ as the \emph{sphere}.
As usual, $\P^1$ has a description in terms of projective coordinates $x=\proj{X}{Y}$,
\be
\P^1(\K) = \{ \proj{X}{Y} \} = 
\bigl\{ (X,Y) \in \K\times\K \setminus (0,0) \bigr\} \,/\, 
\bigl\{ (X,Y) \sim (\lambda X,\lambda Y) \;\;\forall\lambda\in\K^\times \bigr\} \,.
\ee
The conformal symmetries of $\P^1(\K)$ are the M\"obius transformations, which act in projective coordinates $\proj{X}{Y}$ as
\be
\proj{X}{Y} \mapsto \proj{aX+bY}{cX+dY} \,,
\qquad
a,b,c,d\in\K,\quad
ad-bc\ne 0 \,.
\ee
The usual affine coordinates are obtained by setting $Y=1$. 
Two M\"obius transformations whose coefficients $a,b,c,d$ differ by a scalar multiple act equivalently on $\P^1(\K)$, implying that the distinct M\"obius transformations form the group $\PGL(2,\K)$. 
Note that unlike the complex case, $\PGL(2,\K)$ is not the same as $\PSL(2,\K)$, since not every number has a square root in $\K$.

To avoid clutter, in what follows the norm $|\cdot |$ will always refer to the norm $|\cdot |_\mathbb{K}$ on $\K$ unless otherwise stated.

\subsection{\texorpdfstring{$p$}{p}-adic CFT}
\label{sec:pcft}
In this paper we work with the axiomatization of $p$-adic CFT by Melzer \cite{Melzer}.
We emphasize that these axioms are not the most general ones that could be considered;
for example, his axioms do not allow for local operators carrying spin, even though $p$-adic AdS/CFT has been generalized to incorporate a version of spin \cite{Gubser:2018cha}. 

We thus take a $p$-adic CFT on $\P^1(\K)$ to be defined by the following data:
\begin{enumerate}[nolistsep,label*=(\arabic*)]

\item A list of primary fields $\cO_a$ ($a\in\cP$) and their conformal dimensions $\Delta_a$.
\item Correlation functions
\be
\corr{\cO_{a_1}(x_1)\cdots \cO_{a_m}(x_m)} \in \C ,
\qquad
x_1,\ldots,x_m \in \K 
\ee
that are continuous functions of the $x_i$'s and independent of the order of insertion.
(We only consider bosonic CFTs.)

\item An operator product expansion: inside correlation functions, the identity
\be
\cO_a(x)\cO_b(y) = \sum_{c\in\cP}C^c{}_{ab}(x,y)\cO_c(y) 
\ee
is satisfied, provided that the distance between $y$ and $x$ is no larger than the distance between $y$ and any other insertion in the correlation function. 
\item $\PGL(2,\K)$ covariance: for $\mu$ a M\"obius transformation,
\be
\corr{(\mu^*\cO_{a_1})(x_1)\cdots(\mu^*\cO_{a_m})(x_m)} =
\corr{\cO_{a_1}(\mu(x_1))\cdots \cO_{a_m}(\mu(x_m))}
\ee
where $(\mu^*\cO_a)(x) = |\frac{ad-bc}{(cx+d)^2}|^{-\Delta_a}\cO_a(x)$.
\end{enumerate} 

\medskip
To be consistent, the CFT data must satisfy \textit{crossing symmetry}, namely the condition that the operator product algebra must be consistent with the structure of correlation functions.
Given an $n$-point correlation function ($n\ge 2$), we can reduce it iteratively to a 1-point function by successively taking the OPE of pairs of nearby operators, yielding an expression for the full $n$-point function.
Crossing symmetry can be stated as the requirement that the result be independent of the order in which we perform allowed contractions.

The main obstacle for solving these constraints in a 2D CFT is that, in general, the OPE coefficients for arbitrary descendants are not known in closed form.
$p$-adic CFT, on the other hand, is particularly simple because there are no descendants. 
This is a consequence of the fact that the operators take values in $\C$ but are continuous functions on $\K$. 
Such functions are locally constant, and therefore all their derivatives vanish~\cite{Koblitz}. 
As a result, the OPE takes the form
\be
\cO_{a}(x_1)\cO_{b}(x_2) = \sum_{c\in\cP}C^c{}_{ab}|x_1-x_2|^{\Delta_c-\Delta_a-\Delta_b} \cO_c(x_2) \,.
\ee
As usual, there must be an identity operator $\cO_1=1$ of dimension $\Delta_1=0$ and whose OPE coefficients satisfy $C^a{}_{b1}=C^a{}_{1b}=\delta^a_b$.
We assume that $1$ is the only operator with this scaling dimension, in which case scale invariance implies that $1$ is the only operator with a 1-point function.
Inserting the OPE into a 2-point function then implies that
\be
\corr{\cO_a(x)\cO_b(y)} = C^1{}_{ab}|x-y|^{-2\Delta_a} \,.
\ee
Furthermore, M\"obius covariance implies as usual that $C^1{}_{ab}=0$ unless $\Delta_a=\Delta_b$. 
For the 3-point function we obtain
\be
\corr{\cO_a(x)\cO_b(y)\cO_c(z)}
=
\frac{C_{abc}}
{|x-y|^{\Delta_{ab}} |y-z|^{\Delta_{bc}} |z-x|^{\Delta_{ca}}} \,,
\ee
\be
C_{abc} = \sum_d C^1{}_{ad}C^{d}{}_{bc} \,,
\qquad
\Delta_{ab} = \Delta_a + \Delta_b - \Delta_c \text{, etc.}
\ee
Order independence of the correlator means that $C_{abc}$ is (for a bosonic CFT) permutation symmetric. 
In particular, the algebra defined by $C^a{}_{bc}$ is abelian: $C^a{}_{bc}=C^a{}_{cb}$. 
When computing 3-point function with insertions at $x$, $y$ and $z$, one encounters the important feature that two of the coordinate distances always coincide: for example, if $|x-y|<|x-z|\le|y-z|$, then $|x-z|=|y-z|$. 

Turning now to the 4-point function, a field at infinity is defined by
\be
\cO_a(\infty) = \lim_{|x|\to\infty} |x|^{2\Delta_a} \cO_a(x) \,.
\ee
A general 4-point function with cross ratio $\eta$ can be written in terms of the correlator
\be
\corr{ \cO_{a}(\infty) \cO_{b}(1) \cO_{c}(\eta) \cO_{d}(0) }
= \left\{ \begin{array}{lll}
\displaystyle
\sum_e C_{abe}C^e{}_{cd} |\eta|^{\Delta_e-\Delta_{c}-\Delta_{d}}
& \quad & |\eta|\le|1-\eta|=1 \\
\displaystyle
\sum_e C_{ade}C^e{}_{bc} 
	|1-\eta|^{\Delta_e-\Delta_{b}-\Delta_{c}}
	& \quad & |1-\eta|\le|\eta|=1 \\
\displaystyle
\sum_e C_{ace}C^e{}_{bd} 
	|\eta|^{\Delta_{a}-\Delta_{c}-\Delta_{e}}
	& \quad & 1\le|\eta|=|1-\eta| 
\end{array} \right\} \,.
\label{eq:CFT correlators}
\ee
These equations imply (for $\K\ne\Q_2$) that $C^a{}_{bc}$ satisfies the crossing constraint
\be
\sum_{e}C_{abe}C^e{}_{cd}
=\sum_e C_{ace}C^e{}_{bd} = \sum_e C_{ade}C^e{}_{bc} \,,
\ee
which is equivalent to the associativity of the algebra defined by $C^a{}_{bc}$.
We assume these constraints are satisfied even for $\K=\Q_2$.

\subsection{\texorpdfstring{$p$}{p}-adic AdS/CFT}
In vacuum $\AdS_3/\CFT_2$, the points of the dual AdS geometry are in one-to-one correspondence with the maximal compact subgroups of the conformal group. 
Each such subgroup is conjugate to $\PSU(2) \subset \PSL_2(\C)$, and is the subgroup fixing the corresponding bulk point.
$p$-adic $\AdS/\CFT$ begins by assuming the same structure:
the bulk points are in one-to-one correspondence with the maximal compact subgroups of
the conformal group $\PGL_2(\K)$. 
Bulk points can be uniquely linked in such a way that the resulting object is a $(q+1)$-valent tree whose edges and vertices are acted on transitively by $\PGL_2(\K)$.
This object is the Bruhat-Tits (BT) tree.
It was shown in \cite{Gubser:2016guj} that a bulk action principle can be introduced on the BT tree, which can then be used to define an AdS/CFT-type dictionary whose correlation functions behave like those of a $p$-adic CFT.

We will now briefly review the geometry of the BT tree and the $p$-adic holographic dictionary associated to it.

\subsubsection{Bruhat-Tits tree}
Recent reviews of the BT tree can be found in \cite{Gubser:2016guj, Heydeman:2016ldy, Bhattacharyya:2017aly}, so here we limit ourselves to those features most relevant to the current discussion. 

As with Archimedean CFT, the points of the $p$-adic bulk geometry are taken to be in one-to-one correspondence with the maximal compact subgroups of the conformal group $\PGL_2(\K)$. 
These subgroups are conjugates of $\PGL_2(\Z_\K)$, where $\Z_\K=\{x\in\K \,:\, |x| \le 1\}$ is the ring of $p$-adic integers. 
Bulk points can be related to equivalence classes of $\Z_\K$-lattice in $\K^2$, and the $p$-adic valuation can be used to define an integer ``distance'' between these lattices. 
(The detailed construction of the tree in terms of equivalence classes of lattices and the action of $\mathrm{PGL}_2(\Q_p)$ on the tree is reviewed in detail in \cite{Bhattacharyya:2017aly}, and will not be repeated here.)
By linking bulk points separated by distance 1, one obtains the Bruhat-Tits tree: a $(q+1)$-valent tree whose edges and vertices are acted on transitively by $\PGL_2(\K)$. 
We may write the points of the BT tree simply as the quotient
\begin{align}
  \H_\K &= \PGL_2(\K) / \PGL_2(\Z_\K) \,.
\end{align}
Despite the similarities of this construction to the Archimedean case, one crucial difference between $\mathbb{H}_\K$ and $\mathbb{H}$ is of course that the former is discrete. 

\begin{figure}[!ht]
	\centering
	\includegraphics[width=0.6\textwidth]{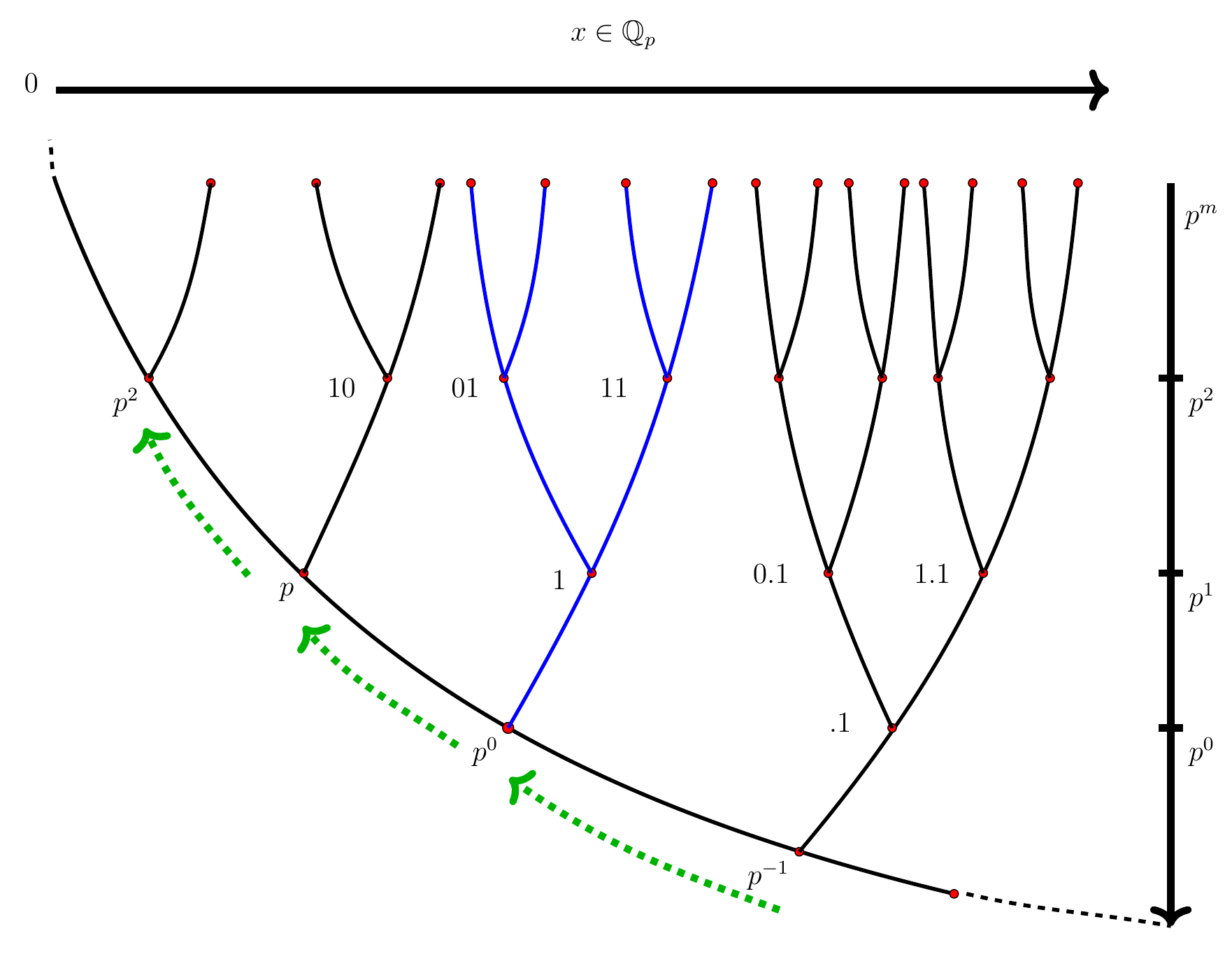}
	\caption{The 2-adic BT tree.}
	\label{fig:padictree}
\end{figure}

There is a convenient choice of parametrization of the tree,  by pairs $(z,x) \in p^\Z \times \K$, see e.g.\ \cite{Brekke:1993gf, Gubser:2016guj}.
This parametrization is illustrated in \figref{fig:padictree} , with the $z = p^v$ coordinate denoting the depth in the ``radial'' direction, and $x \in \K$ describing the ``horizontal'' position. 
Since $p^\Z \subset \K$, we have $|z|=p^{-v}$.
In this description, $z$ tells us the accuracy with which we must specify $x$, meaning the parametrization is not unique: $(z,x) = (z,x')$ whenever $|x-x'| \le |z|$. 

\subsubsection{Bulk action principle}
After identifying the BT tree as the holographic dual geometry to $\Q_p$, \cite{Gubser:2016guj} proposed a $p$-adic version of the standard AdS/CFT dictionary. 
The generating function of the $p$-adic CFT is identified with the path-integral of a dual theory living on the BT tree. 
The path-integral in the bulk is defined in terms of a classical local action for a set of scalar fields.
The generalization to higher spin fields propagating on the tree has been considered recently in \cite{Gubser:2018cha}, but we restrict ourselves to scalar fields for simplicity.

\medskip
Consider a single real-valued scalar field $\phi$ living on $\H_\K$.
Its action is defined in terms of the standard graph Laplacian and takes the form
\be
S= S_2 + S_\text{int},
\ee
where
\be
S_2(\phi,J) = \sum_{\corr{u,v}}\tfrac{1}{2}(\phi(u)-\phi(v))^2
 + \sum_{u}\bigl( \tfrac{1}{2}m^2\phi(u)^2 - J(u)\phi(u) \bigr) 
\ee
is the quadratic part of the action, and $S_\text{int}$ denotes a generic local interaction
\be
  S_\text{int} 
  = S_3 + S_4 +\cdots, \qquad \textrm{with}\qquad S_k = \frac{\lambda_k}{k!} \sum_u \phi(u)^k \,.
\ee

The equation of motion following from $S_2$ is given by
\be
(\Box+m^2)\phi(u) = J(u) \,,
\ee
where the tree Laplacian is
\be
\Box\,\phi(u) = \sum_{\langle uv \rangle} (\phi(u) - \phi(v)) \,.
\ee
The corresponding bulk-to-bulk propagator is given by \cite{Gubser:2016guj, Heydeman:2016ldy}
\be
  G(v;w) = \frac{\zeta_p (2\Delta)}{p^{\Delta}} p^{-\Delta d(v,w)} ,  
  \qquad 
  \qquad m^2 = -\frac{1}{\zeta_p(\Delta-n)\zeta_p(-\Delta)} \,,
\ee
where  $\zeta_p(s) = (1-p^{-s})^{-1}$, $d(v,w)$ is the length of the path between the boundary point $w$ and the bulk point $v$, and $n$ is the degree of $\K$ over $\Q_p$.

The bulk-to-boundary propagator can be defined similarly, with the caveat that the distance between the boundary and any bulk point is divergent, and must be regularized. 
A particular choice of regularization combined with symmetry considerations led to the expression \cite{Gubser:2016guj, Heydeman:2016ldy}
\be
K(z,x;y) = \frac{\zeta_p(2\Delta)}{\zeta_p(2\Delta-n)} \frac{|z_0|^{\Delta}}{|(z, x-y)|^{2\Delta}},
\label{eq:bulk-boundary propagator}
\ee
with 
\be
  |(z, x-y)| = \textrm{sup}\{|z|, |x-y|\}.
\ee
The tree-level boundary 2-point function can be extracted from this expression by regularizing the bulk path integral appropriately. 

Higher-point functions are computed via Witten diagrams. 
It has been shown that they generate 2-, 3-, and 4-point correlation functions with the form expected of a $p$-adic CFT \cite{Gubser:2016guj, Heydeman:2016ldy}. 
They can also be reproduced by PGL$(2,\mathbb{K})$ Wilson line networks living on BT tree \cite{Hung:2018mcn}.

\subsection{The wavefunction approach}
\begin{figure}
        \centering
        \includegraphics[width=0.5\textwidth]{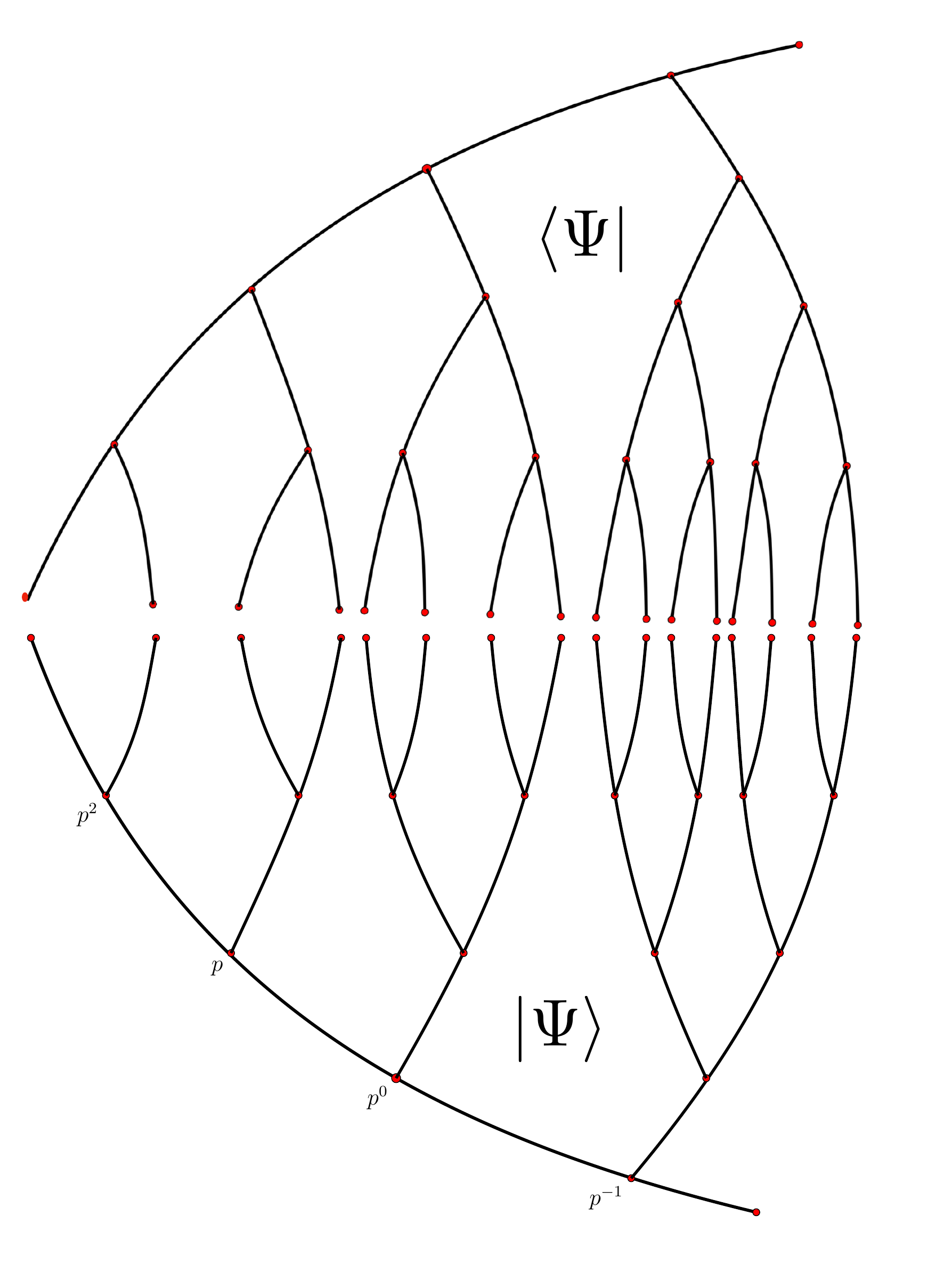}
        \caption{The tensor network of \protect\cite{Bhattacharyya:2017aly} is interpreted as the wavefunction of $p$-adic CFT. Correlation functions are computed by gluing the tensor network and its conjugate along their cutoff surfaces.}
        \label{fig:treecutoff}
\end{figure}

In our previous study relating tensor networks to $p$-adic AdS/CFT \cite{Bhattacharyya:2017aly}, the tensor network plays the role of a ``wavefunction'' for the $p$-adic CFT. 
To compute correlation functions, one glues a tree tensor network with its conjugate along the common boundary, which is the analogue of the Poincar\'e cutoff surface in the BT tree (see \figref{fig:treecutoff}).
Correlation functions thus depend quadratically on the wavefunction.

At each vertex in the tree lives a $(q+1)$-valent tensor $T^{(q+1)}$.
To reproduce a given $p$-adic CFT with OPE coefficients $C_{abc}$, one needs to solve the relation: 
\be \label{non-linear}
  T_{i_1 i_2 \cdots i_{q+1}} 
  \cO_{a_1}^{i_1 \bar{\imath}_1} \cO_{a_2}^{i_2 \bar \imath_2} 
  \cdots \cO_{a_{q+1}}^{i_{q+1} \bar\imath_{q+1}} T^*_{\bar{\imath}_1 \bar{\imath}_2  \cdots \bar{\imath}_{q+1}} = C_{a_1\cdots a_{q+1}} p^{-\Delta_1 -\Delta_2 - ... -\Delta_{q+1}},
\ee
where $C_{a_1\cdots a_{q+1}}$ is defined by
\be 
  C_{a_1\cdots a_{q+1}} 
  = C_{a_1a_2b_1} C^{b_1}{}_{a_3b_2} \cdots C^{b_{q-2}}{}_{a_{q} a_{q+1}}.
\ee
The equation  (\ref{non-linear}) is a set of non-linear algebraic relations involving many parameters, and is difficult to solve analytically. 
This makes writing down a tensor network configuration reproducing the correlators of any particular $p$-adic CFT a challenging problem. 

In the next section, we will show that we can use tensor network, not as a wavefunction,  but as a linear map encoding the Euclidean path integral of any $p$-adic CFT. 
The constraint equation analogous to \eqref{non-linear} is linear in the tensor rather than quadratic, and is possible to solve explicitly.
This makes it straightforward to write down the tensor network configuration corresponding to the path integral of any $p$-adic CFT.

\section{\texorpdfstring{$p$}{p}-adic CFT as a tensor network}
\label{sec:pCFT=TN}
The last section briefly reviewed the $p$-adic AdS/CFT correspondence proposal, wherein the structures of $p$-adic CFT correlators were shown to arise naturally from considering fields living on the BT tree.
In the current section, we strengthen the relationship between $p$-adic CFT and its holographic dual geometry considerably, by showing that $p$-adic CFT is equivalent to a tensor network living on the BT tree.

\subsection{\texorpdfstring{$p$}{p}-adic CFT tensor networks on the Bruhat-Tits tree}
\label{sec:p-adic tensor network}
Our starting point is motivated by a combination of the operator-state correspondence and basic properties of the AdS/CFT correspondence.
Radial quantization of a Euclidean CFT on $\R^n$ associates to any ball $B$ a physical Hilbert space $\cH$ living on its boundary $\p B$. 
The operator-state correspondence states that we may formally associate $\cH$ with the space of local operators acting at the center of the ball. 
General correlation functions may be computed by applying the OPE repeatedly within such a ball, and then taking the overlap $\braket{1}{\Psi}$ of the corresponding state $\ket\Psi$ with the vacuum. 
In the gravitational dual, we may take the state associated to such a ball to live on the minimal bulk surface whose boundary is $\p B$, and we expect the space of gravitational states living on this surface to coincide with that of radial quantization.%
\footnote{Of course, in a gravitational theory the surface need not be minimal, but the minimal surface is always a valid one. There are bad surfaces whose state space is presumably larger than the CFT state space, such as surfaces with non-trivial topology in the bulk.}

Let us apply this intuition to the BT tree.
A $p$-adic ball $B$ consists of all points lying above some bulk vertex $(z,x)$, and we associate to $\p B$ a Hilbert space $\cH$ spanned by the set of local operators. 
We now wish to associate the same state space to the bulk minimal surface ending on $\p B$; in other words, we wish to separate the ball from its complement by cutting the smallest number of bulk legs possible, and associate $\cH$ with the cut legs. 
Since the bulk geometry is a tree, the minimal surface cuts precisely one leg (\figref{fig:cut}). 
\begin{figure}[!ht]
  \begin{subfigure}[t]{0.6\textwidth}
	\centering
    \ifdefined\pTN\else
  \documentclass[tikz]{standalone}
  \usepackage{tikz,etoolbox,xstring}
  \usetikzlibrary{calc,positioning,fit,shapes.misc}
  \begin{document}
\fi

\def\unit		{1cm}
\def\yskip		{1.2cm}

\begin{tikzpicture}[level distance=\yskip]

\tikzstyle{every node}=[draw=none]

\tikzstyle{level 4}=[every node/.style={draw=none,fill=none},sibling distance=\unit/2]
\tikzstyle{level 3}=[sibling distance=\unit]
\tikzstyle{level 2}=[sibling distance=2*\unit]
\tikzstyle{level 1}=[sibling distance=4.2*\unit,parent anchor=center,child anchor=center,edge from parent/.style={draw,thick,black}]

\coordinate (root) at (4.7*\unit,0cm) {} [grow'=up]
 child  foreach \A in {0,1} { 
   coordinate (\A) {} 
   child foreach \B in {0,1} { 
     coordinate (\A\B) {}
     child foreach \C in {0,1} { 
       coordinate (\A\B\C) {} 
       child foreach \D in {0,1} { 
         coordinate (\A\B\C\D) {} 
       }
     }
   }
 };


\draw[blue,very thick] (root) -- (0);
\draw[red,very thick] (0) -- (00);
\draw[blue,very thick] (00) -- (000) -- (0000);
\draw[blue,dotted,very thick] (root) -- ++(3*\unit,-\unit/2) node[shape=coordinate] (infty) {};

\coordinate (ca) at ($(0000)-(0.2\unit,0)$);
\coordinate (ca1) at ($(000)-(0.2\unit,0)$);
\coordinate (cb) at ($(00)-(0,0.2\unit)$);
\coordinate (cc1) at ($(001)+(0.2\unit,0)$);
\coordinate (cc) at ($(0011)+(0.2\unit,0)$);
\draw[dashed] plot [smooth] coordinates {(ca) (ca1) (cb) (cc1) (cc)};

\node [above] at (ca) {$\partial B$};
\node [below] at (cb) {$\mathcal{H}$};
\node [above] at (cc) {$\partial B$};

\end{tikzpicture}

\ifdefined\pTN\else
  \end{document}
\fi
	\caption{}
  \end{subfigure}
  \begin{subfigure}[t]{0.4\textwidth}
	\centering
    \ifdefined\pTN\else
  \documentclass[tikz]{standalone}
  \usepackage{tikz,etoolbox,xstring}
  \usetikzlibrary{calc,positioning,fit,shapes.misc}
  \begin{document}
\fi

\begin{tikzpicture}[scale=1.8]
  \tikzstyle{line}=[draw,thick];
  \def\xa{.6}
  \def\xb{\xa+0.6}
  \def\xc{\xb+0.4}
  
  \path[line,blue,thick] (0,-.5) -- (0,0);
  \path[line,red,very thick] (0,0) -- (0,1);
  \path[line,blue,thick] (0,1) -- (0,2.5);
  
  \path[line,dotted,rounded corners=5mm] (-0.15,-0.5) node[below] {$\partial B$} -- (-0.15,0.5) node[above] {$\mathcal{H}$} -- (\xc,0.5) node[right] {$\partial B$};
  
  \foreach \yai in {0,1,2} {
    \def\ya{\yai}
    \path [line] (0,\ya) -- (\xa,\ya);
    
    \foreach \ybi in {-1,1} {
      \def\yb{\ya+0.25*\ybi}
      \path [line] (\xa,\ya) -- (\xb,\yb);
      
      \foreach \yci in {-1,1} {
        \def\yc{\yb+0.15*\yci}
        \path [line] (\xb,\yb) -- (\xc,\yc);
      }
    }
  }
\end{tikzpicture}

\ifdefined\pTN\else
  \end{document}
\fi
	\caption{}
  \end{subfigure}
  \caption{(a) The Hilbert space $\mathcal{H}$ associated to $\partial B$. (b) A representation of the Bruhat-Tits tree adapted to the $p$-adic cylinder, corresponding to radial quantization. The central trunk is the bulk line between the ball's center and $\infty$, while the boundary of each branching from the central trunk is a sphere ($|x|=\text{constant}$).}
  \label{fig:cut}
\end{figure}
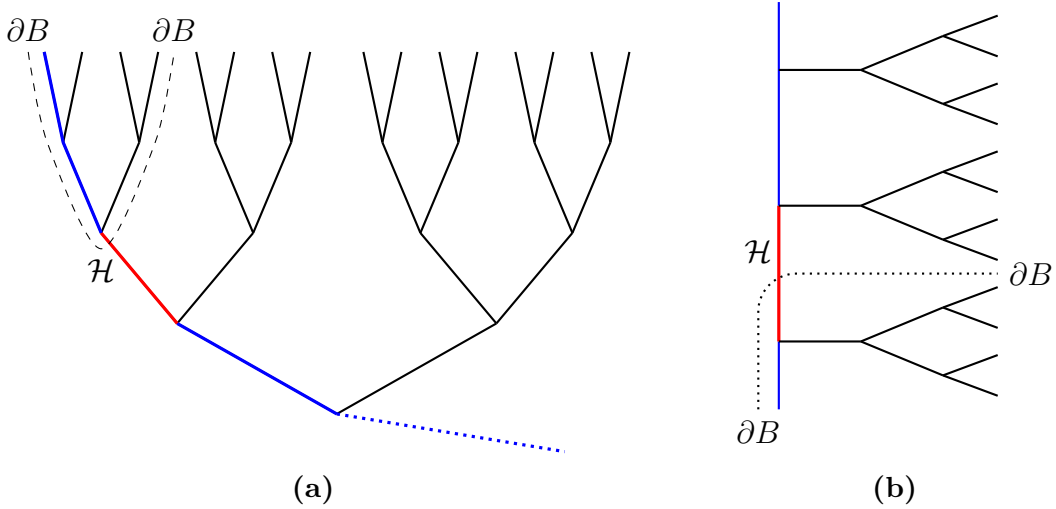
This suggests that to each bulk leg is naturally associated a copy of the space of local operators.

Thus, we introduce on each leg of the BT tree a vector space $\cH$ labeled by the set of local operators.
(Note that in $p$-adic CFT there are no descendants, so all local operators are primary.)
We think of holographic regularization as truncating the Bruhat-Tits tree at some finite level, so that CFT operator insertions are mapped to insertions of states in $\cH$ on the cut edges. 
If we have two nearby operator insertions $\cO_a(x_1)$ and $\cO_b(x_2)$, the OPE tells us we can write this as a sum of insertions of primaries $\cO_c$ at $x_2$. 
After moving the cutoff surface sufficiently far to the infrared, $\cO_a$ and $\cO_b$ will be attached to the same vertex.
This implies that the tensor at the vertex should implement the OPE.

This suggests describing the OPE as an action of $\cH$ on itself as a subalgebra of its own operator algebra, i.e., an injective complex linear map 
\begin{equation}
  \cC : \cH\to\End(\cH)
\end{equation} 
determined by the OPE. 
In particular, $\cH$ should be a $C^*$ algebra.
Let $\{\ket a\}$ be a basis of $\cH$. 
We express $\cC$ in terms of structure constants $C{}^a{}_{bc}$ of the OPE:
\be
\cC : \ket{c} \mapsto \sum_{a,b} \ket{a}C^a{}_{cb} \bra{b} 
\,.
\ee
To reproduce the structure of the OPE, we require $(\cC,\cH)$ to satisfy the following properties:
\begin{enumerate}
\item \emph{Commutativity:} $\cC(a)\cC(b)=\cC(b)\cC(a)$ for all $a,b\in\cH$. 
Equivalently, $\cC^c{}_{ab}$ is symmetric in $a$ and $b$.
	\label{commutativity}
\item \emph{Conjugation:} There is an anti-unitary map $\imath:\cH\to\cH$ satisfying $\imath^2=\id_\cH$ such that $\cC(\imath(a))=\cC(a)^\dagger$ for all $\ket{a}\in\cH$.
	\label{conjugation}
\item \emph{Identity:} $\cH$ has a distinguished element $\ket{1}$ such that $\cC(1)=1_{\cH}$, $\braket{1}{1}=1$, and $\imath\ket{1}=\ket{1}$.
	\label{identity}
\end{enumerate}
Property~\ref{conjugation} is motivated by the fact that we identify $\cH$ with a subspace of $\End(\cH)$, the space of operators on $\cH$. 
The inner product on $\cH$ defines Hermitian conjugation on operators, so it is natural to require this operation to pull back to a complex conjugation on $\cH$ itself.
We denote $\imath\ket{a}=\ket{\bar a}$.
In CFT language, this is the map sending a field $\phi(x)$ to its Hermitian conjugate $\phi^\dagger(x)$.

We wish $\cC$ to behave as $p$-adic OPE coefficients, so we further require:
\begin{enumerate}
\item[4.]
\emph{Associativity:}
$\cC$ satisfies the crossing relation
\be\label{eq:associativity}
\sum_e C^a{}_{be} C^e{}_{cd} = \sum_e C^a{}_{ce} C^e{}_{bd} \,.
\ee
\end{enumerate}
Defining 
\begin{equation}
  C_{abc} = \sum_e C^1{}_{ae} C^e{}_{bc} \,,
\end{equation}
it is straightforward to verify that properties~\ref{identity} and \ref{commutativity} imply $C_{abc}$ is totally symmetric. 
The object 
\begin{equation}
  C_{ab} := C^1{}_{ab} = \bra{1}\cC(a)\cC(b)\ket{1} = \braket{\bar a}{b} = \braket{\bar b}{a}
\end{equation}
is a symmetric non-degenerate bilinear form, which plays the role of a metric on the space of operators. 
We will use it and its inverse $C^{ab}$ to lower and raise indices.

We have yet to introduce operator scaling dimensions.
They appear in an object we call the local bulk propagator.
\begin{enumerate}
\item[5.] \emph{Propagator:} There is a symmetric bivector $G\in\cH\otimes\cH$ satisfying $G(\bra{1},\bra{1})=1$.
\end{enumerate}
The components in our basis are 
\begin{equation}
G^{ab} = G(\bra{a},\bra{b}) \,.
\end{equation}
Defining the operator $\cG:\cH\to\cH$ by 
\begin{equation}
  \bra{a}\cG\ket{b} = \cG^a{}_b = G^{ac} C_{cb} \,,
\end{equation} 
we can construct the scaling dimension operator
\be
\Delta = -\log_p \cG \,.
\ee
In the analogue of a unitary CFT we would require $\cG$ to be Hermitian, so that all operator dimensions are real, but such a restriction is optional.
The standard operator basis is one in which $C_{ab} = \delta_{ab}$ and $G^{ab}$ is diagonal;
the symmetry of $G$ guarantees that such a basis always exists.
In this case, $\cG^a{}_b$ takes the natural form 
\begin{equation}
  \cG^a{}_b = \delta^a_b p^{-\Delta_a} \,,
\end{equation}
where $\Delta\ket{a} = \Delta_a\ket{a}$.
In much of what follows, we implicitly choose such an eigenbasis of $\Delta$, although we do not assume $C_{ab}=\delta_{ab}$. 
Note the crucial property that $C_{ab}=0$ unless $\Delta_a=\Delta_b$. 

\begin{figure}
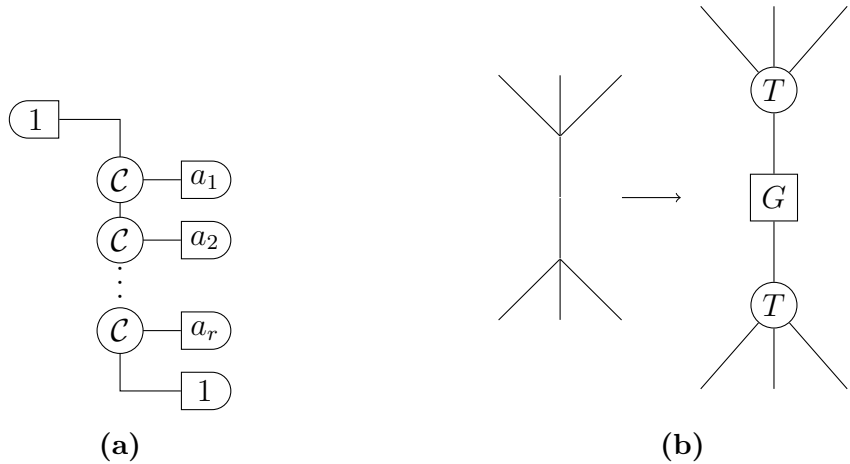

\hspace{\stretch{1}}
\begin{subfigure}[t]{0.4\textwidth}
	\centering
	\ifdefined\pTN\else
  \documentclass{article}
  \usepackage{tikz,amssymb}
  \usetikzlibrary{calc,quotes,positioning,shapes.misc}
  \input{ket.tex}
  \newcommand\cC{\mathcal{C}}
  \begin{document}
\fi

\begin{tikzpicture}[node distance=8mm and 8mm]
 \def\rcirc{.25}

 \begin{scope}[local bounding box=tensor]
  \coordinate                   (c1);
  \coordinate[below=of c1]      (c2);
  \coordinate[below=12mm of c2] (cn);
  \coordinate[above left=of c1] (p0);
  \coordinate[right=of c1]      (p1);
  \coordinate[right=of c2]      (p2);
  \coordinate[right=of cn]      (pn1);
  \coordinate[below right=of cn]     (pn);
  
  \node[bra] (1in)  at (p0)  {$1$};
  \node[ket] (a1)   at (p1)  {$a_1$};
  \node[ket] (a2)   at (p2)  {$a_2$};
  \node[ket] (an1)  at (pn1) {$a_r$};
  \node[ket] (1out) at (pn)  {$1$};
  
  \draw (1in.out) -| (c1) -- (a1.in);
  \draw (c2)      -- (a2.in);
  \draw (cn)      -- (an1.in);
  \draw (cn)      |- (1out.in);
  \draw (c1)      -- (c2);
  \path (c2)      -- node[auto=false, sloped]{$\dots$} (cn);
  
  \node[tensor] at (c1) {$\cC$};
  \node[tensor] at (c2) {$\cC$};
  \node[tensor] at (cn) {$\cC$};
  
 \end{scope}
\end{tikzpicture}

\ifdefined\pTN\else
  \end{document}
\fi
	\caption{}
	\label{fig:tensor}
\end{subfigure}
\hspace{\stretch{1}}
\begin{subfigure}[t]{0.5\textwidth}
	\centering
	\ifdefined\pTN\else
  \documentclass{article}
  \usepackage{tikz}
  \usetikzlibrary{calc,quotes,positioning,shapes.misc}
  \input{ket.tex}
  \begin{document}
\fi

\begin{tikzpicture}[node distance=8mm]
  \begin{scope}[local bounding box=graph]
  \tikzstyle{every node}=[draw,circle,fill=black,minimum size=0pt,
                            inner sep=0pt];
    
    \node (p2) at (0,0) {};
    
    \node[above=of p2] (p1) {};
    \node[above=of p1] (i2) {};
    \node[left=of i2] (i1) {};
    \node[right=of i2] (i3) {};
    
    \node[below=of p2] (p3) {};
    \node[below=of p3] (o2) {};
    \node[left=of o2] (o1) {};
    \node[right=of o2] (o3) {};
    
    \draw (i1) -- (p1) -- (i2);
    \draw (i3) -- (p1) -- (p2) -- (p3) -- (o1);
    \draw (o2) -- (p3) -- (o3);
  \end{scope}  
    
  \begin{scope}[local bounding box=tensor,shift={($(graph.east)+(2,0)$)}]
  \tikzstyle{C}=[draw,circle,fill=none,minimum size=15pt,
                            inner sep=0pt];
    \tikzstyle{P}=[draw,rectangle,fill=none,minimum width=12pt, minimum height=12pt];
    
    \node[propagator] (p2) at (0,0) {$G$};
    
    \node[tensor,above=of p2] (p1) {$T$};
    \node[above=of p1] (i2) {};
    \node[left=of i2] (i1) {};
    \node[right=of i2] (i3) {};
    
    \node[tensor,below=of p2] (p3) {$T$};
    \node[below=of p3] (o2) {};
    \node[left=of o2] (o1) {};
    \node[right=of o2] (o3) {};
    
    \draw (i1) -- (p1) -- (i2);
    \draw (i3) -- (p1) -- (p2) -- (p3) -- (o1);
    \draw (o2) -- (p3) -- (o3);
  \end{scope}
  
  \draw[->] (graph.east) -- (tensor.west);

\end{tikzpicture}

\ifdefined\pTN\else
  \end{document}
\fi
	\caption{}
	\label{fig:network}
\end{subfigure}
\hspace{\stretch{1}}
\caption{(a) Construction of the $r$-valent tensor $T^{(r)}_{a_1\cdots a_r}$ from $\cC^a{}_{bc}$. (b) Tensor network associated to a graph. Here, circles denote the tensor $T^{(4)}_{abcd}$ and the box denotes contraction with $G^{ab}$.}
\end{figure}

\medskip
Now that we have a complete collection of objects encoding our CFT data, we must still attach them to $(q+1)$-valent trees. 
To do this, we define a tensor 
\begin{equation}T^{(r)}\in(\cH^*)^{\otimes r}
\end{equation} for any valence $r$ as follows: with respect to a real orthonormal basis $\{\ket{a}\}$, take
\be \label{eq:fusetensor}
T^{(r)}_{a_1\cdots a_r} \equiv \bra{1}\cC(a_1)\cdots \cC(a_r)\ket{1}
\ee
(\figref{fig:tensor}).
Because the algebraic structure on $\cH$ is commutative, $T^{(r)}$ is totally symmetric.

To any finite graph $\Gamma$ with boundary (that is, hanging edges) $\p\Gamma$ we now associate the following structures, depicted in \figref{fig:network}.
At each vertex $v\in\Gamma$ with valence $r$, we place $T^{(r)}$.
At each internal edge, we contract legs using $G$. 
Finally, at each boundary edge we act with $\cG$. 
The total result is an element
\be
\cT_\Gamma \in (\cH_{\p\Gamma})^* = \bigotimes_{e\in\p\Gamma}(\cH_e)^*,
\ee
which gives a complex linear form on the ``boundary Hilbert space"%
\footnote{We use quotes because this should be understood as a multilinear operator generating imaginary time correlation functions, and not a state in a physical Hilbert space. 
Systems in this paper are analogous to a statistical system, and we make no attempt at a $p$-adic version of the canonical formalism.}
\begin{equation}
\cT_\Gamma : \cH_{\p\Gamma} \to \C
\qquad 
\cH_{\p\Gamma} = \bigotimes_{e\in\p\Gamma}\cH_e \,.
\end{equation}

In particular, these rules associate to any $r$-valent graph $\Gamma$ with $N$ external legs a map $\cH^{\otimes N}\to\C$. 
Our claim is that the application of these rules to the (regulated) Bruhat-Tits tree correctly reproduces all correlation functions of the CFT.
Before proving this statement, we must first describe in detail the regularization procedure, and use it to define the holographic dictionary. 

Before we end the section, we note that if we take a vertex from the Bruhat-Tits tree together with the $q+1$ links attached to it, the tensor assignment described above assigns this subgraph the value $p^{-\Delta_{a_1} - \cdots \Delta_{a_{q+1}}} C_{a_1 \cdots a_{q+1}}$. 
This is in complete agreement with the expectation value of a Wilson line network occupying the same subgraph of the BT tree \cite{Hung:2018mcn}.

\bigskip\noindent
\emph{Example: Ising fusion model.}~~%
For concreteness, let us consider a simple example: the $p$-adic CFT whose structure coefficients $C^a{}_{bc}$ equal the fusion algebra coefficients $N^a{}_{bc}$ of the 2D Ising model.  
We call this the \emph{Ising fusion model}. 
Although its structure constants are not numerically identical, like the 2D Ising model this theory has three primaries: $1$, $\varepsilon$, and $\sigma$, and so a basis for $\cH$ is given by $\{\ket{1},\ket{\epsilon},\ket{\sigma}\}$.
The non-trivial structure constants are given explicitly by 
\begin{align}
\cC(\varepsilon)\ket{\varepsilon} &= \ket 1 &
\cC(\sigma)\ket{\varepsilon} &= \ket\sigma \\
\cC(\varepsilon)\ket{\sigma} &= \ket{\sigma} &
\cC(\sigma)\ket{\sigma} &= \ket 1 + \ket\varepsilon \,.
\end{align}
It is natural to take the above basis to be orthonormal.
Under this assumption, the operators are Hermitian, meaning the basis vectors are fixed under $\imath$. 
It is straightforward to check that these choices are consistent with the axioms.
Finally, we will take $\cG$ to be Hermitian, diagonal on our basis, and to satisfy $\cG\ket{1}=\ket{1}$, but will not otherwise restrict the scaling dimensions $\Delta_\varepsilon$ and $\Delta_\sigma$.

\subsection{The regularized generating functional}
\label{sec:regularizedGF}
As in the usual AdS/CFT correspondence, there exist two natural classes of cutoff surfaces, corresponding to the ``Poincar\'e patch'' (associated to affine $\K$) and the ``global patch'' (associated to projective space $\P^1(\K)$).
Because the ``Poincar\'e patch'' regularization requires a further ``infrared" cutoff to rigorously specify the tensor network, we begin with a cutoff associated to the analog of the global patch radial coordinate.

Choose a point $\mathfrak{o}\in\Gamma$ to be the ``center'' of the Bruhat-Tits tree $\H_\K$, and define the regularized space $\H_\K^\Lambda$ to be the tree whose vertices are those $v\in\Gamma$ with $d(\mathfrak{o},v)<\Lambda$ for an integer $\Lambda$, and containing all edges with at least one endpoint lying in $\H_\K^\Lambda$. 
The number of boundary edges is 
\begin{equation}
N=q^\Lambda(1+\frac{1}{q}) \,,
\end{equation} 
and the resulting tensor network has the same number of slots.
We can think of the tensor network as a map 
\begin{equation}
  \cZ_\Lambda : \cH_\Lambda \to \C \,,
  \qquad
  \cH_\Lambda = \bigotimes_{e\in\p\H^\Lambda_\K} \cH_e \,,
\end{equation} 
and it is this map that we wish to identify with the path integral.
In particular, we claim that as $\Lambda\to\infty$, $\cZ_\Lambda$ encodes all correlation functions.

Let us start with the simplest entry in the dictionary.
Since any correlation function is unchanged if we insert the operator 1, we make the following prescription: at every leg where no state (i.e. local operator) is explicitly inserted, we implicitly insert $\ket 1$. 
For example, $\cH_\Lambda$ contains the distinguished element 
\begin{equation}
  \ket{1}_N \equiv \ket{1}\otimes \ket{1}\otimes \cdots 
  = \bigotimes_{e\in \p\H_\Lambda} \ket{1}_e \,.
\end{equation}
The above prescription implies that the (regularized) partition function on $\P^1(\K)$ is
\be
Z_\Lambda = \cZ_\Lambda(1_\Lambda) \,.
\ee
The tensor network axioms imply that the projective space partition function takes the trivial value $Z_\Lambda = 1$ for any $\Lambda$.
We will see in \secref{sec:higher genus} that the partition function is more interesting at higher genus.

More generally, consider perturbing the partition function by the operator $\ket a$ at a boundary point $X$,%
\footnote{Here, $X$ is an abstract label of the point, as opposed to a precise coordinate. We will use small letters such as $z,x$ for the coordinates.}
accomplished by inserting the value 
\begin{equation}
\ket{1}_X+\lambda^{(a)}_X\ket{a}_X
\end{equation}
in the slot $\cH_X$, where $\lambda^{(a)}_X\in\C$.
Taking the derivative with respect to $\lambda^{(a)}_X$ generates an insertion of $\ket a$ at $X$, so $\lambda^{(a)}_X$ is regarded as the source for $a$ at $X$.

The (regularized) generating state is an element of $\cH_\Lambda$ defined by
\be
{\ket J} \equiv \bigotimes_{x\in\p\cT_\Lambda} {\ket J}_X 
\qquad \textrm{with} \qquad
{\ket J}_X \equiv \sum_a J_a(X){\ket a}_X \,,
\ee
in terms of which we may define the regularized generating functional 
\be
Z_\Lambda(J) = \cZ_\Lambda\cdot\ket{J} \,.
\ee
We will prove later in this section that this object encodes all local correlation functions of the CFT.

Although for point-localized sources this expression is simple to deal with, it is harder to work with smeared sources.
This is because for a strongly coupled CFT with smeared sources, it is difficult to remove the cutoff in this expression, because typical entries in $T^{(r)}$ are $O(1)$, giving strong operator mixing at each step. 
We can circumvent this problem in two cases: when the sourced operator is relevant, and when there is a parameter with respect to which all non-trivial $n$-point vertices ($n\ge 3$) are small, as happens in the large-$N$ expansion. 
We will offer minor comments on this point later.
In any event, in those situations where we can define $J$ sensibly while sending $\Lambda\to\infty$, we define the partition function $Z(J)$ to be $\lim_{\Lambda\to\infty} Z_\Lambda(J)$.

In fact, we can give an expression for the general (regularized) correlation function.
Label the external legs by $X_1,\ldots,X_N$.
Then
\be
\corr{\cO_{a_1}(X_1)\cdots \cO_{a_N}(X_N)}_\Lambda
=
\cZ_\Lambda(\ket{a_1}\otimes \ket{a_2}\otimes\cdots\otimes\ket{a_N}) 
\,.
\ee
We can get low-point correlation functions by taking $a_i=1$ for all but a few $i$.
These expressions are obtained by taking functional derivatives of $\cZ_\Lambda(J)$ with respect to the components $J_{a_i}(X_i)$.

\subsubsection*{Poincar\'e patch regularization}
\begin{figure}[h]
	\centering
	\includegraphics[width=0.8\textwidth]{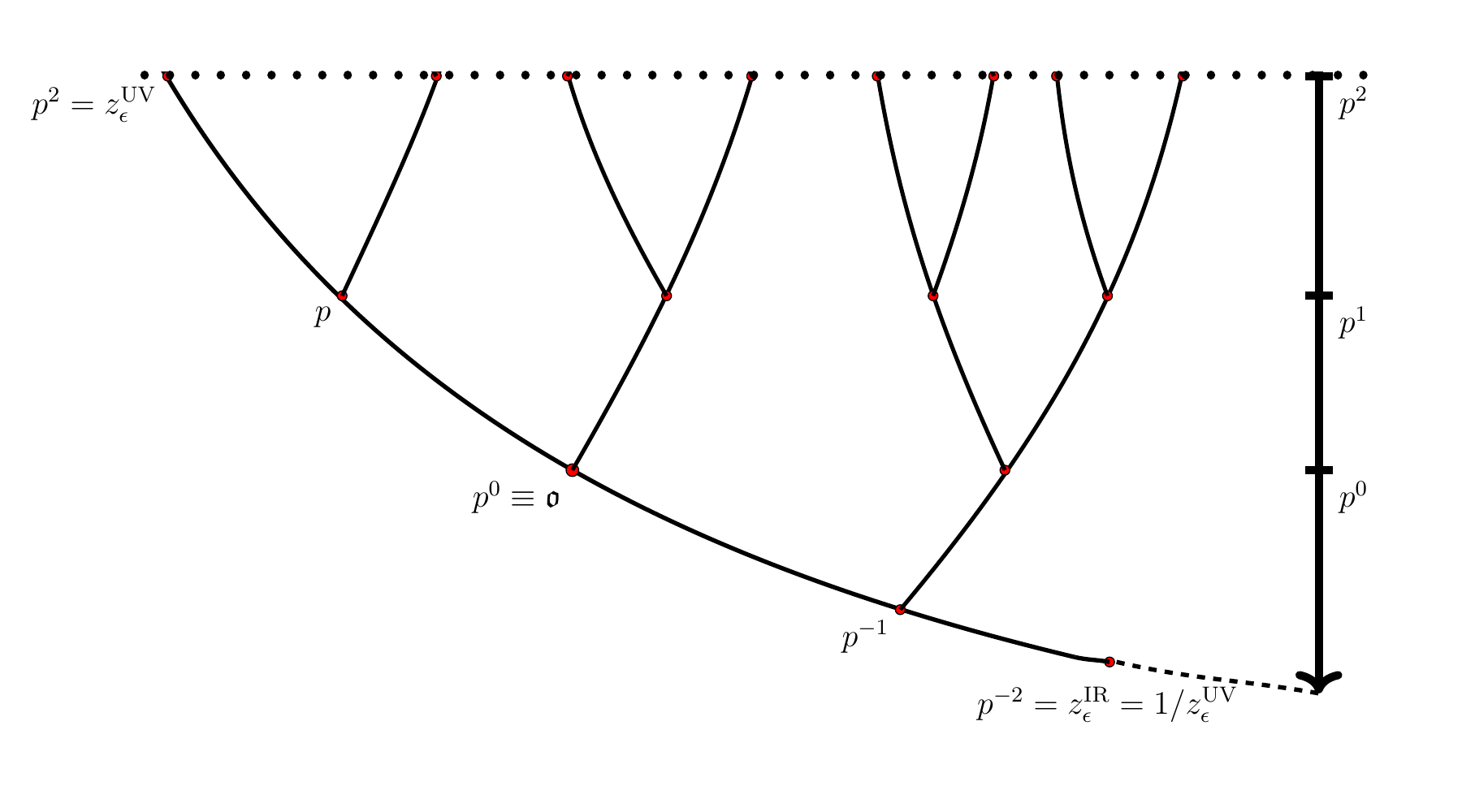}
	\caption{Diagram illustrating the UV and IR cutoff introduced in (\ref{eq:cutoff}). For illustrative purpose, we have picked a UV cutoff at $p^2$, and the IR cutoff at $p^{-2}$.}
	\label{fig:cutoff}
\end{figure}

So far we have worked with the ``global patch'' regularization of $\H_\K$.
Since correlation functions are simplest in the affine ``Poincar\'e patch'' description, however, it is the one we will mostly use from now on.
In this description, our ``radial'' coordinate is given by the level $z$. 
Of course, there are infinitely many points $(z,x)$ at each level, corresponding to the fact that affine space has infinite volume. 
This means we require both a UV and an IR regulator.
While we could choose independent UV and IR regulators, a natural regularization is to fix a cutoff $z_\epsilon$ with small norm $|z_\epsilon|=\epsilon<1$, and consider the set of bulk points
\begin{align} \label{eq:cutoff}
  \H_\K^{(\epsilon)} = \{ (z,x) \;:\: 
  	|z| > \epsilon 
	\quad\text{and}\quad
	|x| < \epsilon^{-1} \} \,,
\end{align}
which is the truncation of the tree spanning $z_\epsilon^{-1}\Z_\K$ to those points with $|z|>\epsilon$. This is illustrated in figure \ref{fig:cutoff}.
This corresponds to choosing $z_\epsilon^\text{IR} = 1/z_\epsilon^\text{UV}$.

The number of hanging edges is $N_\epsilon = q^{-1}\epsilon^{-2n}+1$, giving the regularized Hilbert space $\cH_\epsilon = \cH^{\otimes N_\epsilon}$. 
The final `1' in this expression counts the leg pointing toward `infinity'.

The construction of the regularized generating functional now proceeds as above.
We label each hanging edge by the coordinate $(z,x)$ of its excised vertex; these are all of the form $(z_\epsilon,x)$ excepting the infinity leg, which can be written as $(z_{\epsilon}^{-1},0)$. 
The regularized source vector therefore takes the form
\begin{align}
  \ket{J_\epsilon} &= \bigotimes_{(z,x)\in\p\H_\K^{(\epsilon)}} \ket{J_\epsilon(z,x)}_{(z,x)} \,.
\end{align}

\subsection{Sources and the AdS/CFT dictionary}
In the standard AdS/CFT dictionary, sources are encoded in boundary conditions.
Their values can be recovered from the behavior of the 1-point function of single trace operators,
\be
\corr{\Phi_a(z,x)}_J = z^{n-\Delta_a}J(x) + \cdots \,,
\ee
where $n$ is the dimension of the CFT spacetime.
In this section we study sources on regularized tensor networks, and how to remove the regularization parameter in a controlled way. 

Let us say we wish to add a source $J(x)$ for an operator $\phi$, where now $x$ lives in the $p$-adic field $\K$. 
Our goal is to approximate such a source by an object on the tensor network.
Referring to the above expression, we expect it to be encoded in the 1-point function of bulk operators.
Since the effect of sources is to insert operators in the correlation function, it is natural to realize sources in terms of insertions of this type.
In other words, we expect the source contribution to be schematically of the form
\begin{align}
  e^{\int J(x)\phi(x) dx} \sim \prod_{x\in\K}(1+J(x)\phi(x) + \cdots) \,.
\label{eq:source term}
\end{align}
Let us cut off the Bruhat-Tits tree at level $z=\zeta$, with $\epsilon=|\zeta|$ small.
The boundary points are therefore parametrized by pairs $(\zeta,x)$.

The regularized version of \eqref{eq:source term} involves a regularized current $J_\epsilon(\zeta,x)$ giving the coefficient of the insertion of $\phi$ at $(\zeta,x)$. 
To determine the relation between $J$ and the boundary state, let us insert a state $\ket{1}_{(\zeta,x)} + J_\epsilon(\zeta,x)\ket{\phi}_{(\zeta,x)}$ at each branch and perform one renormalization step toward the infrared, bringing us from level $\zeta$ to level $p^{-1}\zeta$. 
The operator insertion at the point $(p^{-1}\zeta,x)$ is
\begin{align}
\Bigl[ \prod_{(\zeta,y)}(1 + J_\epsilon(\zeta,y) p^{-\Delta}\phi) \Bigr] \ket{1} 
& = \ket{1} + p^{-\Delta} \sum_{(\zeta,y)} J(\zeta,y) \ket{\phi} + O(J^2) 
\end{align}
where the product over $(\zeta,y)$ runs over the $q=p^n$ boundary points lying over $(p^{-1}\zeta,x)$. 
The sum in this expression can be understood as the $p$-adic integral of a function that is locally constant on sets of the form $x+\zeta\Z_\K$, suggesting that the value of $J(\zeta,x)$ should be defined in terms of $p$-adic integration of $J(x)$ over the ball $x+\zeta\Z_\K$.

To be more specific, let us instead try to take the cutoff to infinity without changing the value of $\corr{\phi(\zeta,x)}$ at a fixed bulk point $(\zeta,x)$.
Up to higher order terms, this is accomplished by setting
\begin{align}
  J(\zeta,x) = |\zeta|^{-\Delta} \int_{x+\zeta\Z_\K} \!\!\!\! J(x)\, dx = |\zeta|^{n-\Delta} [J]_{x+\zeta\Z_\K}
\end{align}
where $[J]_S=\frac{1}{\mathrm{vol}(S)}\int_S J(x) dx$ denotes the average value of the function $J$ on a set $S\subset\K$, and we recall that $n$ is the dimension of $\K$ over $\Q_p$. 
Typically we take $J$ to be a continuous, \emph{i.e.} locally constant, function $J:\K\to\C$.
Then the above expression implies that $J(z,x)$ becomes small as $|z|\to 0$ provided $\Delta<n$ --- the standard definition of a relevant operator.

The above considerations assumed that $O(J^2)$ terms could be dropped when $|z|$ is small, but clearly this is only true for relevant operators, or for marginal operators with small sources.
We thereby recover a standard fact of AdS/CFT: sources for relevant operators only affect behavior in the infrared, and UV cutoffs can be removed in their presence.
On the other hand, sources for irrelevant operators lead to drastic modifications of the theory as we try to remove the cutoff, and are not under control from the point of view of the IR CFT. 

The problematic terms for irrelevant operators involve the fusion contributions, for example the $O(J^2)$ term
\begin{align}
  \sum_{(z,y)\ne (z,y')} J(z,y) J(z,y') [\cC(\phi)]^2 \ket{1}_{(z/p,x)} \,.
\end{align}
There are therefore two ways we can still obtain useful information about irrelevant operators.
One is to take $J$ to be a sum of delta functions, corresponding to the computation of a correlation function.
Such an expression involves only a finite number of fusion operations, and remains under control in spite of the divergence in the value of $J(z,x)$ itself.
The second is when the structure coefficients themselves suppress the fusion operation.
This is typical in studies of holography where the semi-classical supergravity approximation is valid. 
In this case, the non-trivial structure coefficients are suppressed by powers of $N$.
While this does not allow one to remove the cutoff entirely, it does allow one to study regularized correlation functions over a wide range of values of $z$.

\smallskip
Note that in evolving the boundary state one step into the interior, we have considered the vertex to be a map $\cH^{\otimes q} \to \cH$, which differs slightly from our definition as a map $\cH^{\otimes(q+1)}\to\C$.
We must therefore convert the leg lying below the vertex, which lives in the dual space $\cH^*$, into a vector.
The correct way to do this is using the inverse $C^{ab}$ of the 2-point coefficient. 
As a consequence the propagator bivector $G$ must be contracted with $C_{ab}$, resulting in the propagator map $\cG$. 
In this way, we can split the tensor network into layers, each of which maps the Hilbert space at one renormalization step to that of the next.
The computations that follow are formulated in this language.

\subsection{Boundary correlation functions}
\label{sec:correlators}
We now turn to the study of the correlation functions of a finite number of local operators. 
An insertion of $\cO(y)$ is accomplished by setting its source to $J(x) = \delta(x-y)$.
The defining property of the delta function,
\begin{align}
  \int_S \delta(x)\,dx = \left\{ \begin{array}{ll}
    1\qquad & 0\in S \,, \\
    0 & \text{otherwise} \,,
  \end{array} \right. 
\end{align}
gives the regularized current
\begin{align}
  J(z,x) &= |z|^{-\Delta} \left\{ \begin{array}{ll}
    1\qquad & |x-y| < |z| \,, \\
    0\qquad & \text{otherwise.} 
  \end{array} \right.
\end{align}
In other words, a single insertion of $\cO_a(x)$ is equivalent at cutoff level $z_\epsilon$ to inserting the state $|z_\epsilon|^{-\Delta}\ket{a}_{(z_\epsilon,x)}$. 

\begin{figure}[b]
	\centering
	\ifdefined\pTN\else
  \documentclass{article}
  \usepackage{tikz}
  \usetikzlibrary{calc,quotes,positioning,shapes.misc}
  \input{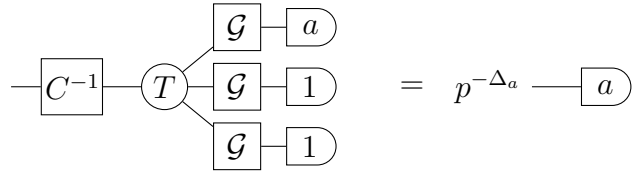}
  \newcommand\cG{\mathcal{G}}
  \begin{document}
\fi

\tikzset{
  cinverse/.style = {vertex, rectangle, draw=black, fill=white, 
                       minimum width=(1.3*\propagatorsize), minimum height=(1.2*\propagatorsize)}
}

\begin{tikzpicture}[node distance=5mm and 5mm]

  \begin{scope}[local bounding box=tensor]
    \node[cinverse] (c) {$\hspace{-.455ex}C^{-1}$};
    \node[tensor,right=of c] (t) {$T$};
    
    \node[right=of t] (p2) {};
    \node[above=of p2] (p1) {};
    \node[below=of p2] (p3) {};
    

    \node[propagator] (g1) at (p1) {$\cG$};
    \node[propagator] (g2) at (p2) {$\cG$};
    \node[propagator] (g3) at (p3) {$\cG$};

    \node[ket,right=of p1] (k1) {$a$};
    \node[ket,right=of p2] (k2) {$1$};
    \node[ket,right=of p3] (k3) {$1$};
    
    \draw (k1.in) -- (g1) -- (t) -- (c) -- ++(-.8,0);
    \draw (k2.in) -- (g2) -- (t) -- (g3) -- (k3.in);
  \end{scope} 
  
  \node at ($(tensor.east)+(1,0)$) {$=$};
  
  \begin{scope}[local bounding box=contracted,shift={($(tensor.east)+(2,0)$)}]
    \node (p1) {$p^{-\Delta_a}$};
    \node[right=of p1] (p2) {};
    
    \node[ket] (k) at (p2) {$a$};
    
    \draw (p1) -- (k.in);
  \end{scope}

\end{tikzpicture}

\ifdefined\pTN\else
  \end{document}
\fi
	\caption{The contribution from the insertion of a single non-trivial operator $a$ on the boundary gets renormalized by $p^{-\Delta_a}$ as we move one step into the interior.}
	\label{fig:renormalization}
\end{figure}

To see this directly in terms of the tensor network, consider the output of a single tensor/propagator pair in the BT tree, with the operator $\ket a$ on one upper leg and $\ket 1$ attached to the other top legs.
The properties of $T^{(r)}$ imply that this is equivalent to the single ket $p^{-\Delta_a}\ket{a}$ (see \figref{fig:renormalization}). 
To keep correlation functions invariant as the cutoff is removed, we must therefore compensate the insertion at each step by a factor $p^{\Delta_a}$, which is accomplished by the $|z|^{-\Delta}$ factor.

At this stage, it is interesting to compute the one-point function of the bulk operator $\cO_a(z,x)$ in the presence of this current.
We can define this bulk operator by adding an extra leg to the interaction vertex $\cT^{(q)}$ located at $(z,x)$, and inserting the state $\ket{a}$ at this leg.
Inserting this into the above prescription for $J(z,x)$ gives
\begin{align}
  \corr{\cO_b(z,x)}_{J_a(x)=\delta(x-y)} &= C_{ab}\left\{ \begin{array}{ll}
    |z|^{-\Delta} & |x-y| \le |z| \,, \\
    \frac{|z|^\Delta}{|x-y|^{2\Delta}} \quad & \text{otherwise,}
  \end{array}\right. 
\end{align}
which we recognize (up to normalization) as the bulk-boundary propagator of equation \eqref{eq:bulk-boundary propagator}.

\subsubsection{2-point function}
We work in the Poincar\'e patch regularization, with the cutoff $z_\epsilon$ chosen such that both operator insertions are contained within the infrared regulator. 
The 2-point function  $\corr{\cO_a(x_1)\cO_b(x_2)}$ is equal to the partition function $Z(J)$, with the current
\begin{align}
  J_a(x) &= \delta(x-x_1) &
  J_b(x) &= \delta(x-x_2) \,.
\end{align}
As described above, the partition function of the regularized network is computed by inserting the states $|z_\epsilon|^{-\Delta_a}\ket{a}$ and $|z_\epsilon|^{-\Delta_b}\ket{b}$ at $(z_\epsilon,x_1)$ and $(z_\epsilon,x_2)$, with $\ket{1}$ inserted at all other legs.

\begin{figure}[h]
	\centering
	\includegraphics[width=0.6\textwidth]{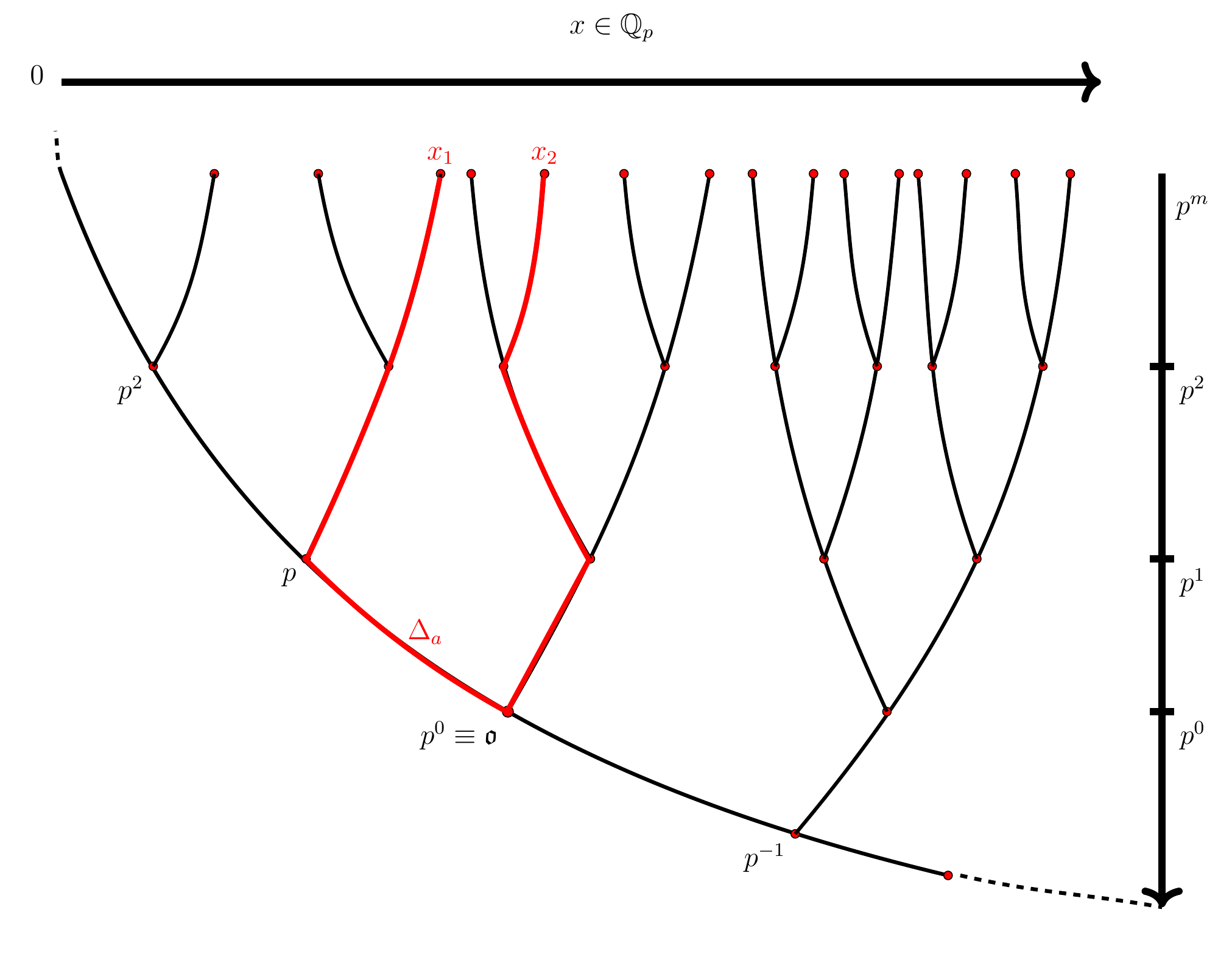}
	\caption{Computation of the 2-point function from the tensor network.}
	\label{fig:2pt}
\end{figure}

When all legs inserted above a vertex are $\ket{1}$, the entire vertex collapses to the insertion of $\ket{1}$ on the leg below it. 
As a result, any branch of the tree disjoint from the path connecting $x_1$ and $x_2$ collapses to $\ket{1}$.
The maximum value of $|z|$ on this path equals $|x_1-x_2|$.
Propagating the boundary states down to the bottom of the geodesic gives $(|z_\epsilon|/|x_1-x_2|)^{\Delta_a+\Delta_b}$, and fusion gives a factor of $C_{ab}$. 
Recalling that $C_{ab}$ and $\Delta$ are simultaneously diagonalizable, we can combine this with the regularization factor to obtain 
\begin{equation}
  \corr{\cO_a(x_1)\cO_b(x_2)} = \frac{C_{ab}}{|x_1-x_2|^{2\Delta_a}} 
\end{equation}
for \emph{any} value of $z_\epsilon$, provided $\epsilon$ is small enough that $(z_\epsilon,x_1)$ and $(z_\epsilon,x_2)$ are distinct points.
This procedure is illustrated in \figref{fig:2pt}.

The computation of the 2-point function is very similar to that for the wavefunction tensor network performed in \cite{Bhattacharyya:2017aly}. 
The interpretation is somewhat different, however, as we are no longer restricted to operator insertion within the same ``spatial slice''.

\subsubsection{3-point function}
\begin{figure}[!h]
	\centering
	\includegraphics[width=0.6\textwidth]{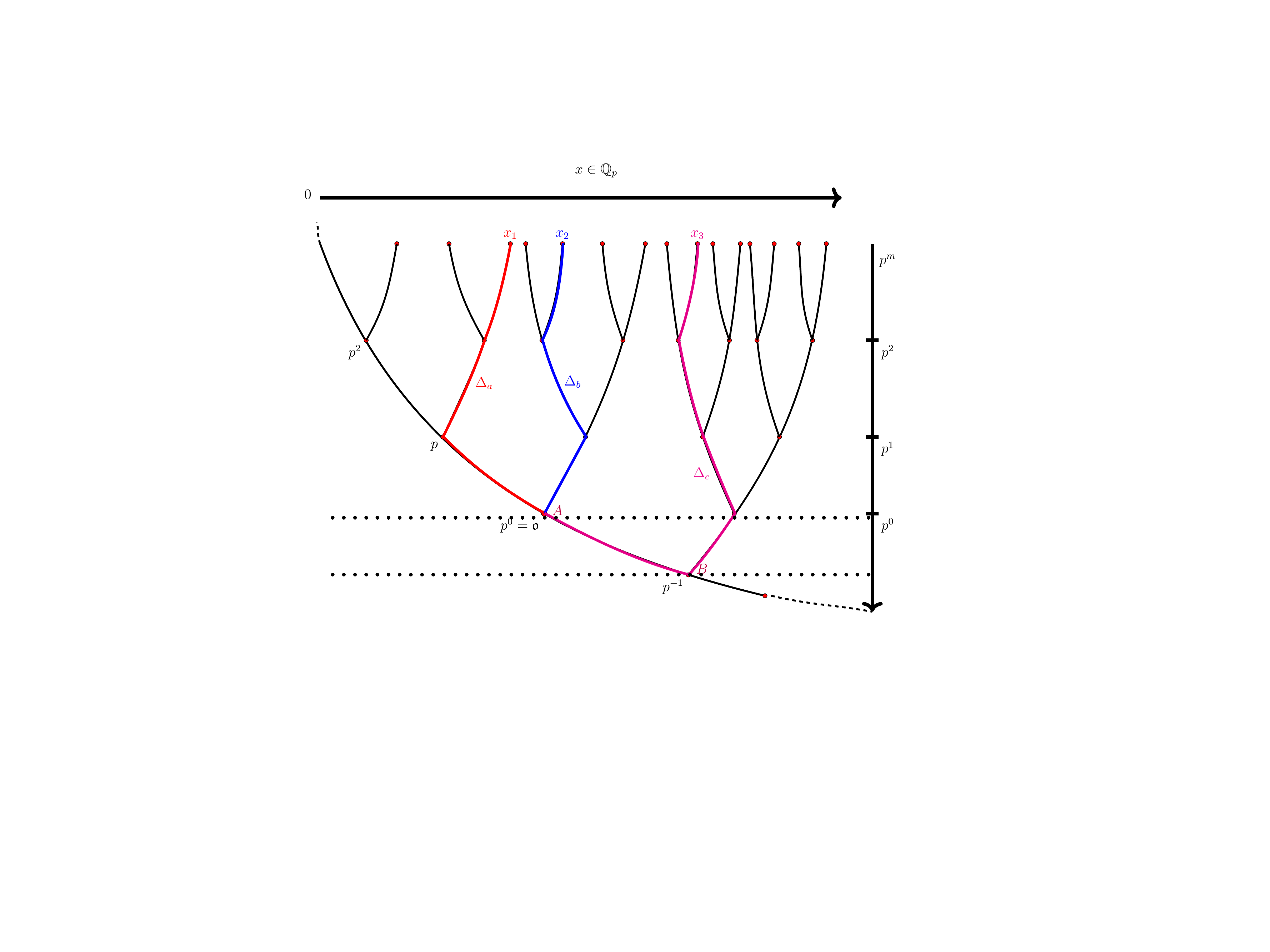}
	\caption{Computation of the 3-point function from the tensor network.\label{fig:3pt}}
\end{figure}

The computation of the 3-point function is quite similar.
Consider the insertion of three operators $\mathcal{O}_a(x_1), \mathcal{O}_b(x_2)$ and $\mathcal{O}_c(x_3)$ at the boundary. 
In a $p$-adic field, any triangle is isosceles with legs longer than the base, and so we may choose our labels so that 
\begin{equation}\label{wlog3}
  |x_{12}| \leq |x_{13}| = |x_{23}| \,.
\end{equation} 
This is illustrated in \figref{fig:3pt}.

Working as before, we apply layers of the tensor network in sequence. 
Non-trivial states propagate along downward-directed paths with weight $p^{-\Delta}$ at each step until two non-trivial states are inserted at a single vertex.
The paths from $x_1$ and $x_2$ meet first at a point we call $A$, where the tensor fuses them into the intermediate operator $\sum_d \ket{d} \cC^d{}_{ab}$.
This operator is propagated until its path intersects that from $x_3$, where they are fused with the 2-point coefficient $C_{cd}$.
Combining these elements with the cutoff-dependent normalization factors from the current gives 
\begin{align}
  \corr{\cO_a(x_1)\cO_b(x_2)\cO_c(x_3)}
  &= \sum_d C_{cd} C^d{}_{ab}|z_\epsilon|^{-\Delta_a-\Delta_b-\Delta_c}
    \left(\frac{|z_\epsilon|}{|x_{12}|}\right)^{\Delta_a+\Delta_b+\Delta_c}
    \left(\frac{|x_{12}|}{|x_{32}|}\right)^{\Delta_d+\Delta_c} \\
  &= \frac{C_{abc}}{|x_{12}|^{\Delta_a+\Delta_b-\Delta_c} |x_{23}|^{\Delta_{b}+ \Delta_c - \Delta_a}|x_{13}|^{\Delta_a+\Delta_c - \Delta_b}} \,,
\end{align}
where the second equality follows from \eqref{wlog3} together with the fact that $C_{cd}$ is non-zero only if $\Delta_c=\Delta_d$.
This confirms that $C_{abc}$ are in fact the CFT structure constants.

Here we would like to pause and contrast our procedure with that of the $p$-adic AdS/CFT story of \cite{Heydeman:2016ldy}. 
There, as in standard AdS/CFT, the correlator is built from bulk-boundary propagators from the insertion points to a bulk interaction vertex, and the position of the vertex has to be integrated over to obtain the final result. 
It is only in the semi-classical limit that the position of the vertex becomes fixed, and the propagators can be replaced by the action of a geodesic stretching from the boundary to the fixed interaction vertex.
In the tensor network formalism, on the other hand, the interaction vertex is located at a fixed position, and no integration over its position is necessary.

\subsubsection{4-point function and onwards}
\begin{figure}[!h]
	\centering
	\includegraphics[width=0.6\textwidth]{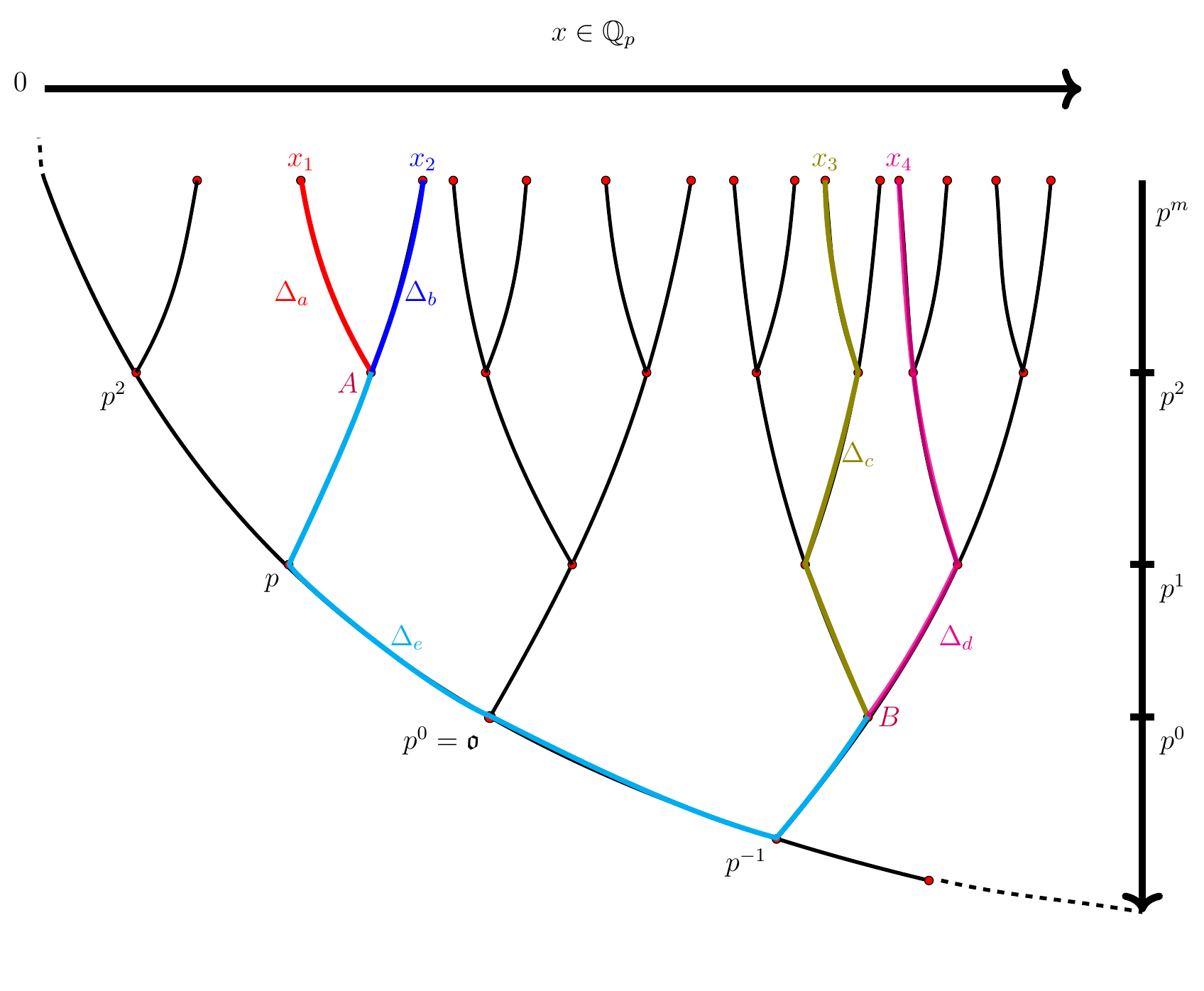}
	\caption{Computation of the 4-point function from the tensor network.
	\label{fig:4pt}}
\end{figure}

Let us now turn to the 4-point function. 
Any four $p$-adic numbers $x_i$ ($i=1,\ldots,4$) have two possible nesting patterns for the six coordinate differences $x_{ij}$. 
With an appropriate labeling, these differences satisfy one of two inequalities:
\begin{align}
|x_{12}| & \leq |x_{13}|=|x_{23}| \leq |x_{14}|=|x_{24}|=|x_{34}|  \quad\text{or} \\
|x_{12}| & \,,\;\: |x_{34}| \leq |x_{13}| =|x_{14}|=|x_{23}|=|x_{24}| \,.
\label{uchannel} 
\end{align}
The two cases can be exchanged by an appropriate $p$-adic M\"obius transformation.
In \cite{Hung:2018mcn} we focused on the first case, so in this paper we work with the second situation, as illustrated in \figref{fig:4pt}.

In the configuration (\ref{uchannel}), the paths from $x_1$ and $x_2$ meet at $A$, and those between $x_3$ and $x_4$ meet at $B$. 
After fusion at the interaction vertices, the result is propagated and fused with $C_{ef}$.
Working in analogy with the 2- and 3-point functions, 
we thus obtain the 4-point function 
\begin{multline}
\corr{ \cO_{a}(x_1) \cO_{b}(x_2) \cO_{c}(x_3) \cO_{d}(x_4) }  \\
\begin{aligned}
&=  \sum_{e,f} C_{ef} C^e{}_{ab} C^f{}_{cd} |x_{12}|^{-\Delta_a - \Delta_b} |x_{34}|^{-\Delta_c - \Delta_d} \Bigl| \frac{x_{12}}{x_{13}} \Bigr|^{\Delta_e} \Bigl| \frac{x_{34}}{x_{13}} \Bigr|^{\Delta_f} \\
&= \sum_e C_{abe}C^e{}_{cd} |x_{12}|^{\Delta_e-\Delta_a-\Delta_b} |x_{34}|^{\Delta_e-\Delta_c-\Delta_d} |x_{13}|^{-2\Delta_e} \,,
\end{aligned}
\label{4pt1}
\end{multline}
where once again we used $\Delta_e=\Delta_f$ whenever $C_{ef}$ is non-zero.
To better understand the result (\ref{4pt1}), we first express it in terms of the conformal cross ratios 
\be
  u = \frac{|x_{12}| |x_{34}|}{|x_{13}| |x_{24}|} 
  \qquad \textrm{and}\qquad   
  v = \frac{|x_{14}| |x_{23}|}{|x_{13}| |x_{24}|} \,,
\ee
giving
\begin{align}
  \sum_e C_{abe}C^e{}_{cd} \frac{u^{\Delta_e}}{|x_{12}|^{\Delta_a+\Delta_b}|x_{34}|^{\Delta_c+\Delta_d}} \,.
\end{align}
Note that $u$ is a natural object in this context: $-\log_p u$ is equal to the distance between the bulk points $A$ and $B$ \cite{Gubser:2017tsi}.
This expression has no dependence on $v$, since $v=1$ when the $x_i$'s satisfy \eqref{uchannel}. 

To obtain a non-trivial $v$ dependence we go to the $v$ channel, which is obtained by exchanging $x_2 \leftrightarrow x_4$ in the relation \eqref{uchannel}. 
The correlator then becomes
\begin{align}
  \corr{ \cO_{a}(x_1) \cO_{b}(x_2) \cO_{c}(x_3) \cO_{d}(x_4) }
  &= \sum_e C_{ade}C^e{}_{bc} \frac{v^{\Delta_e}}{|x_{14}|^{\Delta_a+\Delta_d} |x_{23}|^{\Delta_b+\Delta_c}} \,.
\end{align}

So far, we have only considered situations in which the diagrams involve 3-point vertices alone. 
This suffices for $\Q_2$, where all vertices are 3-valent.
When $q>2$, however, we can consider correlators in which three points are all equidistant from each other: 
\be \label{sit2}
  |x_{12}|= |x_{23}|= |x_{13}| \,.
\ee 
In this case the correlator takes the form
\begin{align}
  \corr{ \cO_{a}(x_1) \cO_{b}(x_2) \cO_{c}(x_3) \cO_{d}(x_4) }
  &= C_{abcd} |x_{12}|^{\Delta_d-\Delta_a-\Delta_b-\Delta_c} |x_{14}|^{-2\Delta_d} \,.
\end{align}
This is a degenerate configuration in which $u=v=1$, and locality suggests that we should require the expressions in both the $u$ and $v$ channels to reduce to this expression when $u=v=1$. 
Imposing this condition results in the relations
\be 
  C_{abcd} = \sum_e C_{abe}C^e{}_{cd} = \sum_e C_{ade}C^e{}_{bc} \,,
\ee
implying the associativity constraints \eqref{eq:associativity}.
The analogous condition for a general $k$-point function is that it can be broken down into repeated products of the structure constants:
\be \label{fusion_chain}
C_{a_1\cdots a_k} = \sum_{\{e_i\}} C_{a_1a_2e_1} C^{e_1}{}_{a_3e_2} \cdots C^{e_{k-3}}{}_{a_{k-1}a_k}.
\ee
Given that the $C_{abc}$ are associative, the $k$-point fusion admits many different equivalent representations in terms of products of $C_{abc}$. 
We thereby arrive at the condition that the tensors $T_{a_1\cdots a_{q+1}}$ should take the form prescribed in (\ref{eq:fusetensor}).

\section{\texorpdfstring{$p$}{p}-adic CFT at higher genus}
\label{sec:higher genus}
Any local 2D CFT has consistent correlation functions, not only on the Riemann sphere $\P^1(\C)$, but also on any Riemann surface $\Sigma_n$ of genus $n$.
In fact, all information about a 2D CFT can be extracted from the variation of its partition function over the moduli space of Riemann surfaces.
It is therefore natural to expect the $p$-adic analogue of this object to be of similar importance in $p$-adic CFT.

One of the simplest ways of working with higher genus curves over $\C$ is the Schottky construction. 
In this approach, one fixes a subgroup $\cS_{\C}\subset\SL(2,\C)$ that is finitely generated and such that every non-trivial element is hyperbolic.%
\footnote{Here by hyperbolic we mean that its eigenvalues have different absolute values, a property which over $\C$ is often termed `loxodromic'.}
Standard theorems tell us that $\cS_{\C}$ is freely generated by some number of generators $n$, known as the rank of $\cS_{\C}$. 
We say $z\in\P^1(\C)$ is a fixed point of $\cS_{\C}$ if $\gamma(z)=z$ for some non-trivial $\gamma\in\cS_{\C}$;
any hyperbolic element has exactly two fixed points.
If $P_\cS$ is the closure of the set of fixed points of $\cS_{\C}$, then $\cS_{\C}$ acts freely and discontinuously on the complement
\begin{equation}
  \Omega_\cS\equiv \P^1(\C)\setminus P_\cS
\end{equation} 
Therefore the quotient
\begin{equation}
  \Sigma^{\mathbb{C}}_\cS \equiv \Omega_\cS/\cS_{\C}
\end{equation} is a compact complex manifold of genus $n$.

Mumford \cite{Mumford} showed that, if we replace $\C$ in this definition by a $p$-adic field $\K$, the resulting $\K$-analytic manifold is an algebraic curve over $\K$.
This manifold has a simple visualization in terms of the BT tree.
Now the Schottky group is a subgroup of the $p$-adic modular group:
\begin{equation}
  \textrm{$p$-adic Schottky group:}\qquad \qquad \mathcal{S}\in \mathrm{PGL}(2,\mathbb{K})
\end{equation}
For any non-trivial $\gamma\in\cS$, there is a unique curve $\cC_\gamma$ in $\cT$ connecting its fixed points.
Let $\Delta_\cS$ be the minimal tree containing all these $\cC_\gamma$.
Again, $\cS$ acts freely and discontinuously on $\Delta_\cS$, allowing us to define the quotient graph 
\begin{equation}
  \textrm{Mumford curve:}\qquad \qquad\Sigma_{\cS}\equiv\Delta_\cS/\cS
\end{equation}
It is a finite graph of genus $n$.
The graph representing $\Sigma_\cS$ is the unique $(p^{n}+1)$-valent topological graph that contracts to $\Sigma$, where $n$ is the degree of the field extension of $\K$ over $\Q_p$.

This graph in hand, it is now straightforward to define $p$-adic CFTs over the Mumford curves by attaching to the graph $\Gamma=\Sigma_\cS$ the tensor network $\cT_\Gamma$.
Certain qualitative aspects of $p$-adic CFT on higher genus curves descending from such graphs were considered from a holographic perspective already in~\cite{Heydeman:2016ldy}.
Assigning a tensor network to this graph gives an explicit construction of the partition function and correlation functions.
It follows immediately that the correlation functions so obtained are identical to those obtained using ``method of images'' procedures to construct correlation functions directly within CFT.
In this section, we discuss the properties of these objects on general Mumford curves.

\subsection{Genus one}
\label{sec:genus 1}

\subsubsection{Genus-one partition function}
A complex genus-one curve corresponds to a choice of modular transformation $\gamma\in\SL(2,\C)$ of infinite order. 
Any such element has two fixed points $z_{1,2}$ on the Riemann sphere, and the curve is given by the quotient $(\P^1(\C)\setminus\{z_1,z_2\})/\langle \gamma\rangle$.
The most familiar parametrization is obtained on mapping $z_{1,2}$ to $\{0,\infty\}$ by modular transformation. 
The action of $\gamma$ is now equivalent to multiplication by a modular parameter $\q\in\C^\times$, which we may take to satisfy $|\q|<1$.

The same process can be repeated in the case of a $p$-adic CFT.
Consider
\begin{equation}
  \K^\times\simeq\P^1(\K)\setminus\{0,\infty\} \,.
\end{equation}
The element
\begin{equation}
  \q\in\K^\times \qquad \textrm{and} \qquad |\q| < 1
\end{equation}
acts on $\K^\times$ by multiplication, generating a group $\cS_{\q}$.
The manifold 
\begin{equation}
  E_{\q}(\K)=\K^\times/\cS_{\q}
\end{equation} 
is then an elliptic curve.
To describe its dual geometry, note that all non-trivial elements of $\cS_{\q}$ share the same endpoints $\{0,\infty\}$.
It follows that the minimal tree $\Delta_\cS$ is simply the bulk geodesic $\cC$ connecting $0$ to $\infty$.
The dual geometry therefore contracts to $\cC/\cS$, which is a loop containing 
\begin{equation}
  k = -\log_p|\q|
\end{equation} 
vertices.
The full bulk graph $\Gamma$ is obtained by attaching infinite trees to each vertex such that the full graph is $(q+1)$-valent.

The simplest observable we can compute is the partition function.
We defined this as the tensor network $\cT_\Gamma$ (or more precisely, any appropriate regularization) with $\ket{1}$ inserted on all boundary legs.
Any tree-like structure with $\ket{1}$ inserted collapses, and we are left with the tensor network $\cT_0$ associated to the loop $\cC/\cS$ of length $k$.
It is simple to evaluate:
\be
\cZ_{\cT_0} = \sum_a \bra{a} \cG^k \ket{a} = \sum_{a} |\q|^{\Delta_a} \,.
\ee
Note that this structure is very similar to the partition function of a complex CFT with diagonal modular invariant,
\be
\cZ = \sum_{\phi} |\q|_\C^{\Delta_\phi-\frac{c}{12}} \sum_{m,\bar m}N(m,\bar m)\q^m \bar{\q}^{\bar m} \,,
\ee
where $N(m,\bar m)$ is the number of descendants at level $(m,\bar m)$.
To obtain the $p$-adic result, we merely replace $\C$ by $\K$, set $c=0$, and drop the descendants from the sum.

\subsubsection{Correlation functions at genus one}
It is also interesting to consider more general correlation functions. 
Correlation functions require specifying a Weyl frame; the natural one for us (and the $p$-adic analogue of choosing a metric of constant curvature on a Riemann surface) corresponds to cutting off at fixed distance from the loop $\cC/\cS$. 
Having done these, we can compute a few classes of correlation functions.
\begin{figure}
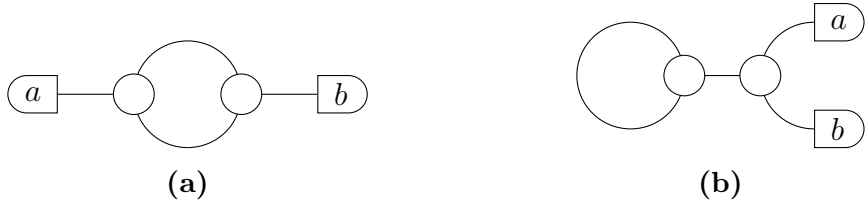

   \centering
   \subcaptionbox{\label{fig:2-pt(a)}}[.4\linewidth]
     {\ifdefined\pTN\else
  \documentclass{article}
  \usepackage{tikz}
  \usetikzlibrary{calc,quotes,positioning,shapes.misc}
  \input{ket.tex}
  \begin{document}
\fi

\begin{tikzpicture}[node distance=10mm]

  \coordinate                    (p1);
  \coordinate[above right=of p1] (a1);
  \coordinate[below right=of p1] (a2);
  \coordinate[below right=of a1] (p2);
  \coordinate[left       =of p1] (pa);
  \coordinate[right      =of p2] (pb);
  
  \node[bra] (a) at (pa) {$a$};
  \node[ket] (b) at (pb) {$b$};
  
  \path
    (a.out) edge               (p1)
  	(p1)    edge [out= 90,in=180]  (a1)
	(a1)    edge [out=  0,in=90]   (p2)
	(p1)    edge [out=270,in=180]  (a2)
    (a2)    edge [out=  0,in=270]  (p2)
    (p2)    edge                   (b.in);
  
  \node[tensor] at (p1) {};
  \node[tensor] at (p2) {};
  
\end{tikzpicture}

\ifdefined\pTN\else
  \end{document}
\fi}
   \qquad
   \subcaptionbox{\label{fig:2-pt(b)}}[.4\linewidth]
     {\ifdefined\pTN\else
  \documentclass{article}
  \usepackage{tikz}
  \usetikzlibrary{calc,quotes,positioning,shapes.misc}
  \input{ket.tex}
  \begin{document}
\fi

\begin{tikzpicture}[node distance=10mm]

  \coordinate (a0);
  \coordinate[above right=of a0] (a1) {};
  \coordinate[below right=of a0] (a2) {};
  \coordinate[below right=of a1] (p1);
  \coordinate[right=of p1] (p2);
  \coordinate[above right=of p2] (pa);
  \coordinate[below right=of p2] (pb);
  
  \node[ket] (a) at (pa) {$a$};
  \node[ket] (b) at (pb) {$b$};
  
  \path
  	(a0) edge [out= 90,in=180]  (a1)
	(a1) edge [out=  0,in=90]   (p1)
	(a0) edge [out=270,in=180]  (a2)
    (a2) edge [out=  0,in=270]  (p1)
    (p1) edge                   (p2)
    (p2) edge [out= 90,in=180]  (a.in)
    (p2) edge [out=270,in=180]  (b.in);
	
  \node[tensor] at (p1) {};
  \node[tensor] at (p2) {};
  
\end{tikzpicture}

\ifdefined\pTN\else
  \end{document}
\fi}
  \caption{2-point function in the cases where (a) the geodesic and the torus' minimal curve intersect, and (b) they do not.}
  \label{fig:2-pt}
\end{figure}

One-point functions are evaluated simply:
\begin{equation}
\corr{\cO_a(x)}_\q = \sum_b \bra{b} \cC(a) \cG^k\ket{b} 
= \sum_b |\q|^{\Delta_b} C^b{}_{ab} \,.
\end{equation}
Due to permutation symmetry of $C_{abc}$, we see that $\cO_a$ can have non-vanishing one-point function only if it lies in the OPE of some operator with itself.
In the Ising fusion model, for example,  
\begin{equation}
\corr{\varepsilon}_\q = |\q|^{\Delta_\sigma} 
\qquad \textrm{and}\qquad
\corr{\sigma}_\q = 0 \,.
\end{equation}

2-point functions $\corr{\cO_a(x)\cO_b(y)}$ can take two forms, depending on $x$ and $y$. 
When the geodesic connecting $x$ and $y$ intersects the curve (\figref{fig:2-pt(a)}), it takes the value
\begin{equation}
  \corr{\cO_a(x)\cO_b(y)} 
  = \sum_{c,d} p^{-\ell\Delta_c - m\Delta_d} C^c{}_{ad}C^{d}{}_{bc}
\end{equation}
where $\ell$ and $m=k-\ell$ denote the two distances between the intersection points of the geodesic connecting $x$ and $y$ with the loop $\cC/\cS$. 
On the other hand, when they do not intersect (\figref{fig:2-pt(b)}) one obtains
\begin{equation}
\corr{\cO_a(x)\cO_b(y)} = \sum_{c,d}|\q|^{\Delta_c}p^{-\ell \Delta_d}C^c{}_{cd}C^{d}{}_{ab} \,, 
\end{equation}
where $\ell$ is the distance from $\cC/\cS$ to the geodesic connecting $x$ and $y$.
The condition that these agree when the geodesics intersect at exactly one point is the associativity constraint \eqref{eq:associativity}.

Note that these are the non-normalized correlation functions; normalized correlators are obtained on division by $\cZ$.

\subsection{Cutting and sewing}
\label{sec:cutting and sewing}
A second useful method of computing correlations on higher genus curves is to use sewing and cutting relations. 
Here we study these relations for $p$-adic CFTs. 
Notably, we will see that the diagrams representing the construction of a curve by cutting and sewing operations is simply the graph $\Delta_\cS/\cS$.

Take local coordinates $z_{1,2}$ on ball-like coordinate patches $U_{1,2}$ of two curves $M_{1,2}$ of genus $n_{1,2}$.
Fix $q_1,q_2\in\K^\times$ and $\beta$ with $0<|\beta|<1$, and excise the disks
$|z_{1,2}| \le |\beta q_{1,2}|$.
We can now sew the curves by identifying 
\begin{equation}
z_1z_2 = q_1q_2
\qquad
\text{for}\quad
|\beta q_{1,2}| < |z_{1,2}| < |\beta^{-1}q_{1,2}| \,.
\end{equation}
The resulting surface has genus 
\begin{equation}
n = n_1 + n_2 + 1
\end{equation}
The details of the surface depend on the choice of local coordinates, as well as the choice of $q=q_1q_2$.
(The individual values of $q_1$ and $q_2$ are irrelevant provided the disks $|z_i|<|\beta^{-1}q_i|$ are contained within the local patches $U_i$.)
If these coordinates are different patches on a single curve, we instead produce a curve of genus $n' = n+1$.

\begin{figure}
  \begin{subfigure}{.7\linewidth}
  \centering
  \ifdefined\pTN\else
  \documentclass{article}
  \usepackage{tikz,etoolbox,xstring}
  \usetikzlibrary{calc,positioning,fit,shapes.misc}
  \begin{document}
\fi

\def\unit		{1cm}
\def\yskip		{1.2cm}

\begin{tikzpicture}[level distance=\yskip]

\def\tickwidth  {2mm}

\tikzstyle{every node}=[draw=black,circle,inner sep=0pt,minimum size=1mm]

\tikzstyle{level 4}=[every node/.style={draw=none,fill=none},sibling distance=\unit/2]
\tikzstyle{level 3}=[sibling distance=\unit]
\tikzstyle{level 2}=[sibling distance=2*\unit]
\tikzstyle{level 1}=[sibling distance=4.2*\unit]

\draw[<->, thick] (0cm,-.5*\yskip) -- (0cm,4.5*\yskip);
\foreach \n in {0,...,4} {
  \pgfmathsetmacro\ppower{int(\n-1)};
  \draw[thick,draw=black] ($(-\tickwidth/2,\n*\yskip)$) -- +(\tickwidth,0)
    node[anchor=west,draw=none,fill=none] {$\scriptstyle p^{\ppower}$};
}

\node (root) at (4.7*\unit,0cm) {} [grow'=up]
 child  foreach \A in {0,1} { 
   node (\A) {} 
   child foreach \B in {0,1} { 
     node (\A\B) {}
     child foreach \C in {0,1} { 
       node (\A\B\C) {} 
       child foreach \D in {0,1} { 
         node (\A\B\C\D) {} 
       }
     }
   }
 };

\draw[green,very thick] (root) -- (0);
\draw[blue,very thick] (0) -- (00);
\draw[black,dotted,very thick] (root) -- ++(3*\unit,-\unit/2) node[shape=coordinate] (infty) {};

\node[rectangle,rounded corners=2mm,draw=red!60!white,fit=(0000)(0011)(00),inner sep=1mm] {};
\node[rectangle,rounded corners=2mm,draw=red!60!white,fit=(1000)(1111)(root)(infty),inner sep=1mm] {};

\draw[<->,red,bend left,shorten <= 3mm, shorten >= 4mm] (0011) edge ($(1000)-(.3*\unit,0)$);

\end{tikzpicture}

\ifdefined\pTN\else
  \end{document}
\fi
  \subcaption{}
  \end{subfigure}%
  \begin{subfigure}{.3\linewidth}
  \centering
  \ifdefined\pTN\else
  \documentclass{article}
  \usepackage{tikz,etoolbox,xstring}
  \usetikzlibrary{calc,positioning,fit,shapes.misc}
  \begin{document}
\fi

\def\unit		{1cm}

\begin{tikzpicture}[level distance=0.7\unit]

\def\tickwidth  {2mm}

\tikzstyle{every node}=[draw=black,circle,inner sep=0pt,minimum size=1mm]

\tikzstyle{level 3}=[every node/.style={draw=none,fill=none},sibling distance=\unit/2]
\tikzstyle{level 2}=[sibling distance=\unit]

\def\radius{.8cm}

\node (r1) {} [grow'=up]
 child {
   node{}
   child  foreach \A in {0,1} { 
     node {} 
     child foreach \B in {0,1} { 
       node {}
     }
   }
 };

\node (r2) at ($(r1)-(0,2*\radius)$) {} [grow'=down]
 child {
   node{}
   child  foreach \A in {0,1} { 
     node {} 
     child foreach \B in {0,1} { 
       node {}
     }
   }
 };

\draw[green,very thick,shorten <=.5mm,shorten >=.5mm] (r1) arc (90:-90:\radius);
\draw[blue,very thick,shorten <=.5mm,shorten >=.5mm] (r1) arc (90:270:\radius);




\end{tikzpicture}

\ifdefined\pTN\else
  \end{document}
\fi
  \subcaption{}
  \end{subfigure}
  \caption{The torus can be constructed by excising a ball around $0$ and $\infty$ and identifying in a toroidal region. This is equivalent to identifying the boxed regions of the BT tree shown in (a). The resulting graph is shown in (b).}
  \label{fig:torus-cut-sew}
\end{figure}
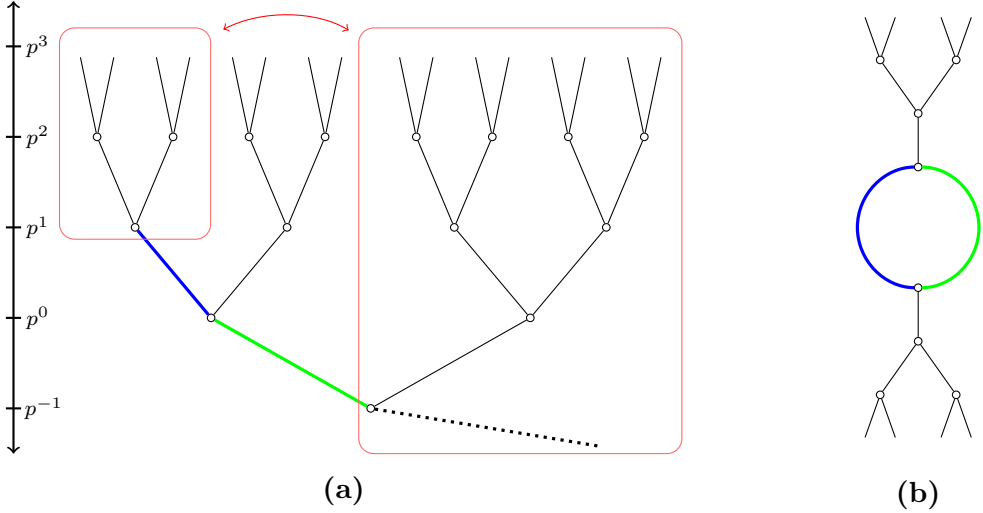

\medskip 
Our primary interest is in the question: what is the resulting bulk graph?
Note that any ball in a coordinate chart has the property that the ball and its complement can be separated by cutting a single bulk edge.
Similarly, any annular region can be obtained by cutting two bulk edges.
The sewing operation described above glues along two annular regions, each containing $-\log_p|\beta|$ vertices.
In fact, the discrete valuation structure gives a simpler way of accomplishing the same operation.
Excise from $U_1$ the region $|z_1|<|q_1|$, and from $U_2$ the region $|z_2|\le |q_2|$ by cutting on a single bulk edge.
(Note that the discrete valuation property implies that both these regions are closed balls.)
Since the overlap region is now empty, no boundary points need to be identified, and gluing is trivial: we simply glue the two cut edges. 

As an example, consider sewing $\P^1$ cut at 0 and $\infty$, shown in figure \ref{fig:torus-cut-sew}.
A natural choice of coordinates are the polar patches, related on the overlap by $z_1z_2=1$.
Choose $q_1$ and $q_2$ satisfying $|q_1|<|q_2^{-1}|$, and glue as above.
In terms of the $z_1$ coordinate system, this corresponds to excising from $\P^1$ the balls $|z_1|<|q_1|$ and $|z_1|\ge|q_2^{-1}|$. 
The bulk graph now has two hanging edges, separated by $-\log_p|q_1q_2|$ vertices.
Gluing hanging edges gives us a $(q+1)$-valent graph, with a single loop containing $-\log_p|q|$ vertices.
This is the same bulk graph obtained from the Schottky construction.

\medskip 
What types of operations can be made on bulk graphs?
The cutting operation was defined in terms of $p$-adic balls, which are diffeomorphic to $\Z_\K$.
The dual geometry of an excised ball is thus a tree with a single cut edge.
This implies that cutting will never sever an edge lying in the skeleton $\Gamma^0$, for the excised graph must always be contractible to a point.
The allowed cuts, therefore, are precisely those of the edges in $\Gamma\setminus\Gamma^0$.
Iteration of cuts and the two sewing operations allow us to construct a unique $p$-adic geometry corresponding to any skeleton graph $\Gamma^0$ with permissible valence.

\subsection{Higher genus}
\label{sec:genus n}
We saw above that a $p$-adic curve $\Sigma$ of genus $g$ corresponds to a $(q+1)$-valent infinite graph $\Gamma$ retracts to a closed finite graph $\Gamma_0$ of genus $g$. 
The covering graph is the BT tree; each quotienting element lives in $\PGL(2,\Q_p)$, so since the fundamental group of the graph is free, the quotient group is a Schottky group.

The corresponding tensor network is now constructed in the obvious way.
To compute correlation functions we must of course regularize. 
A natural regularization is to include every vertex that lies within a distance
$\Lambda$ of $\Gamma_0$, which in the CFT language corresponds to choosing
a uniformized metric on $\Sigma$.

\begin{figure}
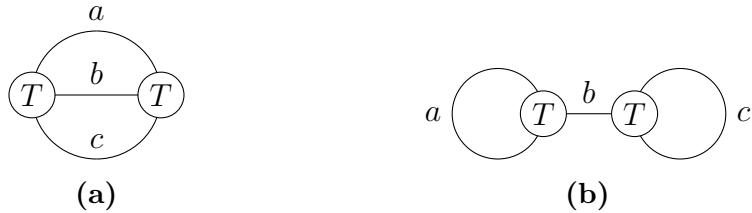

  \centering
  \begin{minipage}[b]{.4\textwidth}
    \centering
    \ifdefined\pTN\else
  \documentclass{article}
  \usepackage{tikz}
  \usetikzlibrary{calc,positioning}
  \input{ket.tex}
  \begin{document}
\fi

\begin{tikzpicture}[node distance=12mm]

  \coordinate (p1);
  \coordinate[above right=of p1,label=above:{$a$}] (a1) {};
  \coordinate[below right=of p1,label=above:{$c$}] (a2) {};
  \coordinate[below right=of a1] (p2);
  
  \path
  	(p1) edge [out= 90,in=180]  (a1)
	(a1) edge [out=  0,in=90]  (p2)
	(p1) edge [out=270,in=180]  (a2)
    (a2) edge [out=  0,in=270]  (p2)
    (p1) edge node[above] {$b$} (p2);
	
  \node[tensor] at (p1) {$T$};
  \node[tensor] at (p2) {$T$};
  
\end{tikzpicture}

\ifdefined\pTN\else
  \end{document}
\fi
    \subcaption{}\label{fig:g=2 gooseberry}
  \end{minipage}\quad%
  \begin{minipage}[b]{.4\textwidth}
   \centering
   \ifdefined\pTN\else
  \documentclass{article}
  \usepackage{tikz}
  \usetikzlibrary{calc,positioning}
  \input{ket.tex}
  \begin{document}
\fi

\def\dumbbellsize{6mm}

\begin{tikzpicture}[node distance=12mm]

  \coordinate (p1);
  \coordinate[left=\dumbbellsize of p1] (c1);
  \coordinate[left=\dumbbellsize of c1] (l1);

  \coordinate[right=of p1] (p2);
  \coordinate[right=\dumbbellsize of p2] (c2);
  \coordinate[right=\dumbbellsize of c2] (l2);
  
  \path (p1) edge node[above] {$b$} (p2);
  \draw (c1) circle (\dumbbellsize)
        (c2) circle (\dumbbellsize);
  
  \node[left]   at (l1) {$a$};
  \node[right]  at (l2) {$c$};
  
  \node[tensor] at (p1) {$T$};
  \node[tensor] at (p2) {$T$};
  
\end{tikzpicture}

\ifdefined\pTN\else
  \end{document}
\fi
    \subcaption{}\label{fig:g=2 dumbbell}
  \end{minipage}
  \caption{The two generic types of $g=2$ graph.}
  \label{fig:g=2}
\end{figure}

We take as an example the partition function, which is computed by the tensor network $\cT_0$ on $\Gamma_0$, and focus on the case where interaction vertices are 3-valent, which is generic in moduli space.
If $\Gamma_0$ has $k$ interaction vertices $X$, we denote by $a_{1,\ldots,3k}$ the Hilbert space index on each leg, and for each edge $e\in E$ we write its pair of $a_i$ indices as $a(e)$.
The partition function is then
\begin{align}
\cZ_{\cT_0} &= \sum_{ \{a\} } C_{a_1a_2a_3}\cdots C_{a_{k-2}a_{k-1}a_k}
\prod_{e\in E} q_e^{\Delta_{a(e)}}\delta_{a(e)} \,,
\end{align}
where $\Delta_{a(e)}$ denotes the dimension of either index (as they are forced to be equal), and $q_e = p^{-\text{length}(e)}$.
For example, for the graph in \figref{fig:g=2 gooseberry},
\begin{align}
  \cZ_{\cT} &= \sum_{a,b,c}(C_{abc})^2 q_a^{\Delta_a} q_b^{\Delta_b} q_c^{\Delta_c} \,.
\end{align}
These expressions coincide with those in a standard 2D CFT, once descendant fields are dropped and geometric terms not relevant to the $p$-adic case have been omitted.

\subsubsection*{Crossing constraints from higher genus partition functions}
Modular invariance plays a crucial role in Archimedean CFT.
For example, the correlators of a CFT can be defined on a general surface precisely when both crossing symmetry and modular invariance on the torus are satisfied.
The property of crossing symmetry may instead be replaced by the condition of modular invariance on higher genus Riemann surfaces, since the crossing relation arises in the computation of higher genus partition functions. 
It is therefore an interesting question to ask: what constraints does the consistency of higher genus partition functions place on a $p$-adic CFT?

Consider the genus 2 partition function.
At a generic point in moduli space, it takes one of the two forms in \figref{fig:g=2}.
It we bring the distance parameter of the $b$ line to zero we reach the degenerate point where $\Delta_b$ is absent from the equation.
If we require the two partition functions to coincide, we obtain the constraint
\begin{equation}
  C^a{}_{ab} C^{bc}{}_c = C^a{}_{bc}C^{bc}{}_a \,,
\end{equation}
which we recognize as a contraction of the associativity constraint \eqref{eq:associativity}. 

If we increase the genus or add operator insertions we can obtain the uncontracted form of the same constraint.
Thus, we can see that crossing symmetry of the OPE coefficients is equivalent to the consistency of the higher genus partition functions.

\section{Renormalization group flow}
\label{sec:rg-flow}
The previous section introduced the machinery of $p$-adic tensor networks, most notably the (regularized) generating functional.
In QFT, the generating functional plays a dual role: in allowing us to generate correlation functions, it also tells us how to deform the theory, perturbing by an operator $\cO$.
In principle, then, the exact generating functional contains information about all RG flows in the theory.

As we have seen, the generating functional of a $p$-adic tensor network has a simple realization as the partition function evaluated on an arbitrary boundary state.
In principle, this allows us to study RG flows exactly.
In practice, a typical RG flow will be difficult to get a precise handle on, as the set of primary operators will be infinite, and operator mixing complex.

Fortunately, the fusion rules of minimal model CFTs give us an infinite set of fusion rules that are simple.
The remainder of this section will be devoted to illustrating the behavior of RG flows in $p$-adic CFTs using these simplest of examples.

\subsection{RG flows and fixed points in \texorpdfstring{$p$}{p}-adic CFT}

\begin{figure}
	\centering
	\ifdefined\pTN\else
  \documentclass{article}
  \usepackage{tikz}
  \usetikzlibrary{calc,quotes,positioning,shapes.misc}
  \input{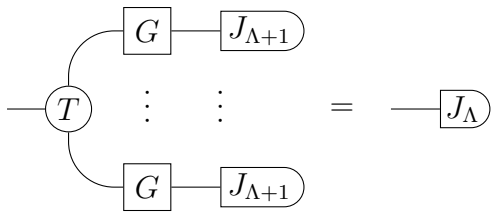}
  \begin{document}
\fi

\begin{tikzpicture}[node distance=5mm and 5mm]

  \begin{scope}[local bounding box=tensor]
    \node[tensor] (t) {$T$};
    \node[left=of t] (p0) {};
    \node[propagator,above right=of t] (g1) {$G$};
    \node[right=of g1] (p1) {};
    \node[propagator,below right=of t] (g2) {$G$};
    \node[right=of g2] (p2) {};
    
    \node[ket] (k1) at (p1) {$J_{\Lambda+1}$};
    \node[ket] (k2) at (p2) {$J_{\Lambda+1}$};
    
    \draw[rounded corners=6mm] (p0) -- (t) |- (g1) -- (k1.in);
    \draw[rounded corners=6mm] (t) |- (g2) -- (k2.in);
    
    \path (g1) -- node[auto=false,sloped]{$\dots$} (g2);
    \path (p1) -- node[auto=false,sloped]{$\dots$} (p2);
  \end{scope} 
  
  \node at ($(tensor.east)+(.5,0)$) {$=$};
  
  \begin{scope}[local bounding box=renormalized,shift={($(tensor.east)+(1,0)$)}]
    \node (p1) {};
    \node[right=of p1] (p2) {};
    
    \node[ket] (rr) at (p2) {$J_\Lambda$};
    
    \draw (p1) -- (rr);
  \end{scope}

\end{tikzpicture}

\ifdefined\pTN\else
  \end{document}
\fi
	\caption{Renormalization of a deformation of a $p$-adic CFT.}
	\label{fig:source renormalization}
\end{figure}

We start with the tensor network of a $p$-adic CFT, whose regularization is parametrized by an integer $\Lambda$ that decreases by 1 during each step to the IR;
for example, in the Poincar\'e patch regularization we could set $z_\epsilon = p^\Lambda$.
Recall that the generating functional takes the form
\be
Z_\Lambda(J_\Lambda) = \cZ_\Lambda\ket{J_\Lambda} \,,
\qquad
\ket{J_\Lambda} = \bigotimes_{X\in\p\Gamma_\Lambda} \ket{J_\Lambda}_X \,,
\quad
\ket{J_\Lambda}_X = \sum_{a\in\cP} J_\Lambda^a(X)\ket{a}_X \,.
\ee
A deformation of a CFT is obtained by replacing the vacuum value $J^a_\Lambda(X)=\delta^a_1$ by an arbitrary $X$-independent value $J^a_\Lambda(X)=J^a_\Lambda$. 
Understanding an RG flow is thus equivalent to understanding the question: how does $\ket{J_\Lambda}$ behave as we change $\Lambda$?
The tensor network construction gives us an explicit answer, which is illustrated in \figref{fig:source renormalization}.
It is convenient to write this in the form
\be
\ket{J_\Lambda}_X = \underbrace{\cC(\cG J_{\Lambda+1})\cdots \cC(\cG J_{\Lambda+1}) }_{p\,\text{times}}
{\ket 1}_X \,.
\label{eq:recursion}
\ee
This equation determines the RG flow of $p$-adic tensor networks.

Correlation functions of operators in the presence of a non-trivial regularized source $J_\Lambda$ are computed as before: varying with respect to $J_\Lambda^a(X)$ still produces a single insertion of ${\ket a}_X$ at the leg $X$. 
The difference is that, while before the state $\ket 1$ was inserted at all other vertices, other vertices now have insertions of the non-trivial state $J_\Lambda$.
To evaluate correlation functions, we must first solve the recurrence relation determining $\ket{J_\Lambda}$.

Fixed points are characterized by the property that $\ket{J_\Lambda}_X$ is independent of $\Lambda$. 
Once we have found a fixed point source $J_*$ of the flow equation, we are faced with the problem of computing correlation functions.
The first item of business is to choose a basis of operators with well-defined scaling dimension.
To do this, we insert an operator at a single leg of the regularized tree and demand that the action of RG flow is to rescale it.
This RG flow action is
\begin{align}
  \ket{\phi_\Lambda} &= [\cC(\cG J)]^{p-1} \cG\ket{\phi_{\Lambda+1}} \,,
\end{align}
from which we obtain the local bulk propagator at the IR fixed point:
\begin{align}
  \cG' = [\cC(\cG J_*)]^{p-1} \cG \,.
\label{eq:IR propagator}
\end{align}
Note that the fixed point equation takes the form $\cG'\ket{J_*} = \ket{J_*}$, so that $\ket{J_*}$ has dimension zero.
$J_*$ plays the role of a ``renormalized identity'' operator, so we will denote it by $\cI$ in what follows.

\bigskip\noindent
\emph{Correlators and structure constants in the IR CFT.}~~%
Given a fixed point of \eqref{eq:recursion}, we now wish to understand the structure of the infrared theory. 
Most importantly, does the infrared theory satisfy the axioms of $p$-adic CFT? 
Answering this question requires understanding the structure coefficients in the IR theory.
While these are determined by the correlation functions, the process of RG flow induces an extra scaling factor into all quantities, which is equal to the sphere partition function.
We must therefore be careful to identify each object properly.

It is instructive to begin with the 2-point function, which after regularization is computed by inserting $\ket{\cI}$ at all but two legs. 
In each renormalization step, vertices with only $\ket{\cI}$s above them are fused to $\ket{\cI}$, while vertices with a single non-trivial insertion $\ket a$ above them fuse to $\cG'\ket a$.
The only non-trivial vertex is the one where the two insertions $\ket a$ and $\ket b$ are fused.
According to our usual axioms, these would first be propagated with $\cG'$ and then fused with the IR coefficients $\hat C'_{ab}$.
However, in this picture they are propagated at the last step with the UV propagator $\cG$ and then fused with the UV coefficients $C_{ab}$. 
We can obtain the unnormalized IR coefficients by requiring these two processes to coincide.
In equation form, this reads
\begin{align}
  \hat C'(\cG'a,\cG'b) 
  &= \bra{1} (\cG a)(\cG b)(\cG\cI)^{p-1} \ket{1} \\
  &= \bra{1} \bigl((\cG\cI)^{p-1}\cG a\bigr) \bigl( (\cG \cI)^{p-1} \cG b\bigr) \bigl( (\cG \cI)^{p-1}\bigr)^{-1} \ket{1}\ \,.
\end{align}
Using \eqref{eq:IR propagator} and the fixed point equation $(\cG \cI)^p=\cI$, we may write%
\footnote{In writing this equation, we have assumed that $\cG \cI$ is an invertible operator. In general this may not be true; if so, such cases can presumably be dealt with by taking ``$(\cG \cI)^{-1}$'' to be the inverse on the subspace perpendicular to the kernel. 
The existence of such a kernel would be quite interesting, because it would mean the IR model has \emph{singular vectors}, giving a clear manifestation of the principle that degrees of freedom are lost in the IR. 
However, as we do not know any explicit models in which $\cG\cI$ is not invertible, this phenomenon may or may not be realized in actual $p$-adic RG flows.}
\begin{align}
  \hat C'(a,b) &= \bra{1} a \cI^{-1} b \ket{\cG \cI} \,.
\end{align}
The normalized 2-point coefficient is found by dividing through by the partition function,
\begin{align}
  C'(a,b) &= \frac{\bra{1} a \cI^{-1} b \ket{\cG \cI}}{\bra{1}\cI\ket{\cG \cI}} \,.
\end{align}

We compute the 3-point coefficients using the same procedure. 
With $\hat C'(a,b,c)=\hat C'_{abc}$, the unnormalized coefficients take the form 
\begin{align}
  \hat C'(\cG'a, \cG'b, \cG'c) &= \bra{1} (\cG a)(\cG b)(\cG c)(\cG \cI)^{p-2} \ket{1} \,,
\end{align}
from which we obtain
\begin{align}
  \hat C'(a,b,c) &= \bra{1} a \cI^{-1} b \cI^{-1} c \ket{\cG \cI} \,.
\end{align}
The (normalized) coefficients $C'^a{}_{bc}$ of the OPE algebra can now be obtained from the relation 
\begin{align}
  \hat C'_{abc} = \sum_d \hat C'_{ad}C'^d{}_{bc} \,.
\end{align}
It is convenient to denote the IR OPE product by $\Pi$, that is, $\Pi(b,c)=\sum_a (C'^a{}_{bc}\,a)$ as an element of $\End(\cH)$.
In state language, the above equation now reads
\begin{align}
  \bra{1} a \cI^{-1} b \cI^{-1} c \ket{\cG \cI} 
  &= \bra{1} a \cI^{-1} \Pi(b,c) \ket{\cG \cI} \,,
\end{align}
from which we obtain the product
\begin{align}
  \Pi(b,c) &= b \cI^{-1} c \,.
\label{eq:IR product}
\end{align}
Equivalently, the IR operator-state mapping $\cC':\cH\to\End(\cH)$ can be written $\cC'(a) = [\cC(\cI)]^{-1}\cC(a)$. 
In retrospect, we could have guessed this from the beginning: it is the unique product law living in the image of $\cC$ satisfying the crucial property of the operator-state mapping that the (renormalized) identity state $\ket \cI$ acts on $\cH$ as the identity operator. 

To satisfy the axioms of $p$-adic CFT, we must prove that $C'^a{}_{bc}$ are commutative and associative, and that the higher-point functions are built from them as described in \secref{sec:p-adic tensor network}. 
Commutativity and associativity are easy to prove, because the IR product $\Pi$ is itself simply the UV product multiplied by $\cC(\cI)^{-1}$:
\begin{align}
  \Pi(a,\Pi(b,c)) &= \Pi(\Pi(a,b),c) = \Pi(a,\Pi(c,b)) = a \cI^{-1} b \cI^{-1} c \,.
\end{align}
Since the product is associative, we may write higher products in the form $\Pi(a_1,a_2,\ldots)$ without ambiguity.

The axioms of $p$-adic CFT also require the higher-point coefficients to be described in terms of these products. 
These arise in correlators when more than 3 geodesic segments intersect at a point. 
As with the 3-point coefficients, the $m$-point coefficient is defined by
\begin{align}
  \hat C'(\cG' a_1,\ldots,\cG' a_m) &= \bra{1} (\cG a_1)\cdots(\cG a_m)(\cG \cI)^{p+1-m}\ket{1} \,,
\end{align}
so that
\begin{align}
  \hat C'(a_1,\ldots,a_m) &= \bra{1} a_1\cdots a_m (\cG \cI)^{1-(m-1)p} \ket{1} \\
  &= \bra{1} a_1 \cI^{-1} a_2 \cdots \cI^{-1} a_m \ket{\cG \cI} \,.
\end{align}
This expression can be written in the form
\begin{align}
  \hat C'(a_1,\ldots,a_m) 
  &= \hat C'(a_1, \Pi(a_2,\ldots,a_m)) \,,
\end{align}
which is precisely the form taken when \eqref{eq:fusetensor} is written in terms of unnormalized coefficients. 
This proves that the IR theory is a $p$-adic CFT.

Finally, let us comment on the structure of the state space itself. 
From the above equations, it is obvious that our axioms in terms of $\cH$ are not satisfied, since $\cI$ does not appear in the same way as the identity. 
The origin of this is that the coefficients are unnormalized.
Recalling the state relation $\braket{\bar a}{b} = C_{ab}$, we see that the Hilbert norm relevant to the IR theory differs from the UV theory. 
Dividing out by the sphere partition function, we can obtain the normalized coefficients 
\begin{align}
  C'(a,b) &= \frac{\bra{1} a \cI^{-1} b \ket{\cG \cI}}{\bra{1}\cI\ket{\cG \cI}} \,.
\end{align}
This implies that the norm required for our state construction is
\begin{align}
  \widetilde{\braket{\bar a}{b}} 
  &= \frac{\bra{\bar a} \cI^{-1}(\cG \cI)\ket{b}}{\bra{1} \cI(\cG \cI)\ket{1}} \,.
\end{align}
In particular, it satisfies $\widetilde{\braket{\bar \cI}{\cI}}=1$ as it must.

\subsection{Example: the Ising fusion model}
It is instructive to consider an explicit example of such an RG flow.
Here, we will study an RG flow in the Ising fusion model of \secref{sec:p-adic tensor network}. 
We deform the theory at cutoff $\Lambda$ by the operator $\varepsilon$, so that
\begin{equation}
  \ket{J_\Lambda} = a_\Lambda\ket{1} + b_\Lambda\ket\varepsilon 
\end{equation}
is inserted on each boundary leg in place of $\ket{1}$.
This form is preserved under renormalization because $1$ and $\varepsilon$ form a closed subalgebra $\cA$ under fusion. 

Denoting the operator dimension of $\varepsilon$ by $\Delta$ to reduce clutter, we may write the fusion of $J$'s in the form
\be
\ket{J_{\Lambda-1}} =
(\cG J_\Lambda)^p \ket 1 = (a_\Lambda \Lambda + p^{-\Delta}b_\Lambda\varepsilon)^p \ket 1 \,.
\ee
Using the fusion identity 
\begin{equation}\varepsilon^n = \frac{1+(-1)^n}{2} + \frac{1-(-1)^n}{2}\varepsilon
\end{equation}
we obtain
\be
a_{\Lambda-1} = \frac{(a_\Lambda + p^{-\Delta}b_\Lambda)^p + (a_\Lambda-p^{-\Delta}b_\Lambda)^p}{2}
\qquad
b_{\Lambda-1} = \frac{(a_\Lambda + p^{-\Delta}b_\Lambda)^p - (a_\Lambda-p^{-\Delta}b_\Lambda)^p}{2} \,.
\ee
Note that an overall rescaling of $\ket J$ is the $p$-adic equivalent of including a counterterm of the form $\int dx\, C$ to the action, and we may eliminate it by dividing the entire partition function by a term of this form.
The physically relevant information is therefore contained in the coupling $\lambda=\frac{b}{a}$, whose recursion relation is
\be
\lambda_{\Lambda-1} = \frac{(1+\lambda_\Lambda p^{-\Delta})^p - (1-\lambda_\Lambda p^{-\Delta})^p}{(1+\lambda_\Lambda p^{-\Delta})^p + (1-\lambda_\Lambda p^{-\Delta})^p} \,.
\ee
An RG fixed point $J=\cI$ is stable under this recursion, being a solution of the equation
\be
\frac{1-\lambda_*}{1+\lambda_*} = \left(\frac{1-\lambda_* p^{-\Delta}}{1+\lambda_*p^{-\Delta}}\right)^{p} \,.
	\label{eq:fixed point}
\ee
Given such a solution, we can find the appropriate value of $a_\Lambda$ such that $a_\Lambda,b_\Lambda$ are stationary at the fixed point:
\be
a_* = \left[ \frac{(1+\lambda_* p^{-\Delta})^p + (1-\lambda_*p^{-\Delta})^p}{2} \right]^{-\frac{1}{p-1}} \,.
\ee

Consider the situation where $\lambda_*$ is small and perturbation theory applies.
Expanding both sides to 2nd order in $\lambda$, we obtain in addition to $\lambda=0$ the solution
\be
\lambda_* = \frac{1}{1+p^{1-\Delta}} + \cdots \,,
\ee
which is a good solution provided $p^{1-\Delta}\gg 1$.
This is true when $\Delta<1$ for any sufficiently large $p$, or with any $p$ for $\Delta$ a sufficiently large negative number.
As an expansion in this parameter, we find the fixed point
\be
\lambda_* \approx \frac{1}{p^{1-\Delta}} + \cdots \,.
\ee
This is in contrast to the situation in the Ising model over $\C$, where there are no nontrivial IR fixed points.

\subsubsection*{Fixed point CFT for $p=2$}
For $p=2$ we can solve \eqref{eq:fixed point} exactly:
\be
\lambda_* = \pm 2^{\Delta}\sqrt{2^{1-\Delta}-1} 
\qquad \textrm{and}\qquad
a_* = 2^{\Delta-1} \,,
\ee
giving two fixed points with real $\lambda_*$ if $\Delta<1$. 
To find the other scaling operator $\phi$ built from $\varepsilon$, we require the action $\cI$ on $\phi$ during a renormalization step to scale it by a constant, i.e. we choose $\phi=\varepsilon+\gamma\,1$ so that 
\begin{equation}
  (\cG \cI)(\cG\phi) 
  = a_*(1+2^{-\Delta}\lambda_*\varepsilon)(2^{\Delta}\varepsilon + \gamma)
  = p^{-\Delta_\phi}\phi 
  \,.
\end{equation}
Equivalently,
\begin{align}
  \gamma 
  & = \frac{\gamma+2^{-2\Delta}\lambda_*}{2^{-\Delta}(\gamma\lambda_*+1)} \,, 
  & 2^{-2\Delta'_\phi}
  & = a_*2^{-\Delta}(\gamma\lambda_*+1) \,.
\label{eq:op recursion}
\end{align}
The non-trivial solution to this equation is
\begin{align}
  \gamma &= -2^{-\Delta}\lambda_* = \mp \sqrt{2^{1-\Delta}-1} \,, &
  \Delta'_\phi 
  &= 1 - \log_2(2^\Delta-1) \,.
\end{align}
Since $\Delta<1$, $\phi$ is irrelevant, which is consistent with the fact that it controls the RG flow near the IR.
Note moreover that $\Delta_\phi$ is complex if $\Delta<0$.

Finally, consider the $\sigma$ field.
Since our deformation preserves the $\Z_2$ Ising symmetry $\sigma\mapsto-\sigma$, we expect it not to mix with other operators, and this is indeed the case.
The IR operator dimension $\Delta_\sigma'$ is related to its UV dimension $\Delta_\sigma$ by
\be
(a_*+b_*2^{-\Delta}\varepsilon)(\sigma 2^{-\Delta_\sigma})
= 2^{-\Delta'_\sigma}\sigma \,,
\ee
giving
\be
\Delta'_\sigma = \Delta_\sigma - \Delta + 1 
- \log_2\bigl[ 1 \pm \sqrt{2^{1-\Delta}-1} \bigr] \,.
\ee

To derive what (if any) $p$-adic CFT is obtained in the IR limit, the natural starting point is to compute correlation functions.
Simplest is the partition function itself, which has $\cI$ inserted at every boundary edge.
Using the relation 
\begin{equation}
  (\cG \cI)^2=\cI \,,
\end{equation}
the tree collapses down to a vertex, leaving the partition function
\be
Z_\Lambda = \bra{1} (\cG \cI)^3 \ket{1} 
= \bra{1} \cI (\cG \cI) \ket{1}
= a_*^2 + 2^{-\Delta}b_*^2 
= 2^{2(\Delta-1)} \bigl[ 3-2^\Delta \bigr] \,.
\ee
We divide this out when evaluating normalized correlation functions. 
It is of interest to note that $0<Z_\Lambda<1$ when $0<\Delta<1$, so that the value of the partition function on $\P^1$ decreases from the UV to the IR.
This may reflect of a more general $F$-theorem--like principle in $p$-adic CFTs.

Consider now the 1-point function $\corr\phi$.
Picking a base point to evaluate the final contraction, up to a power of $2^{-\Delta'_\phi}$ we obtain
\be
\corr\phi \propto \bra{1}(\cG\phi)(\cG \cI)^2\ket{1}
 = \gamma + 2^{-2\Delta}\lambda_*(2+\lambda_*\gamma) 
 = 0 \,,
\ee
as must be the case in a CFT.
Symmetry under $\sigma\to-\sigma$ guarantees $\corr{\sigma}=0$.
The unnormalized 2-point function $\corr{\phi\phi}$ is equal to $\hat C'_{\cI\phi\phi}|x-y|^{-2\Delta'_\phi}$, with
\begin{align}
  \hat C'_{\cI\phi\phi} 
  &= 2^{2\Delta'_\phi}\bra{1}(\cG\phi)^2\ket{\cG\cI} 
   = 2^{1-\Delta}\frac{3-2^\Delta}{2^\Delta-1} \,.
\end{align}
Similarly, the 2-point function $\corr{\sigma\sigma}$ takes the same form, with
\begin{align}
  \hat C'_{\cI\sigma\sigma} &= \frac{2^{1-\Delta}}{1\pm\sqrt{2^{1-\Delta}-1}} \,.
\end{align}
Computation of the 3-point functions proceeds similarly.
Unnormalized 3-point correlators take the standard form
\begin{align}
  \corr{\cO_a(x_1)\cO_b(x_2)\cO_c(x_3)} = \frac{\hat C'_{abc}}{|x_{12}|^{\Delta'_a+\Delta'_b-\Delta'_c}|x_{13}|^{\Delta'_a+\Delta'_c-\Delta'_b}|x_{23}|^{\Delta'_b+\Delta'_c-\Delta'_a}} \,.
\end{align}
Symmetry guarantees that the only non-trivial ones are the $\phi\sigma\sigma$ and $\phi\phi\phi$ correlators, and the unnormalized coefficients are
\begin{align}
  \hat C'_{\phi\sigma\sigma} 
  &= 2^{\Delta'_\phi+2\Delta'\sigma}\bra{1}(\cG\phi)(\cG\sigma)^2\ket{1} 
   = \frac{2^{2(1-\Delta)} (2^\Delta - 1)}{(1 \pm 2^\Delta \sqrt{2^{1-\Delta}-1})^2} \, \\
  \hat C'_{\phi\phi\phi}
  &= 2^{3\Delta'_\phi}\bra{1}(\cG\phi)^3\ket{1} = \mp 2^{3-2\Delta} \frac{(3-2^{\Delta})(2^{\Delta}+1)\sqrt{2^{1-\Delta}-1}}{(2^{\Delta}-1)^3} \,.
\end{align}

Finally, let us verify explicitly that associativity is satisfied.
The only condition that is not trivially satisfied is from the $\phi\sigma\sigma$ channel, and can be written
\begin{align}
  \frac{(\hat C'_{\phi\sigma\sigma})^2}
  {\hat C'_{\cI\sigma\sigma}C'_{\cI\phi\phi}}
  &=
  \frac{\hat C'_{\cI\sigma\sigma}}{C'_{\cI\cI\cI}}
  + \frac{\hat C'_{\phi\phi\phi}\hat C'_{\phi\sigma\sigma}}
  {(\hat C'_{\cI\phi\phi})^2} \,.
\end{align}
This relation is indeed satisfied.

\section{Discussion}
We have shown in this paper that a CFT living on the $p$-adic number line $\Q_p$, as axiomatized in \cite{Melzer}, is equivalent to a tensor network on its holographic dual geometry, the BT tree. 
Since the BT tree is infinite, it must be cut off to regularize it, yielding a finite tensor network with hanging edges. 
Interpreting insertions on the hanging edges as sources, the regularization process becomes directly analogous to the holographic dictionary, and as the cutoff is removed we obtain the generating functional of the $p$-adic CFT. 
The CFT correlation functions extracted from this object satisfy the OPE of the $p$-adic CFT, and therefore reproduce all correlation functions.

We applied our results in two ways.
The first is to consider correlation and partition functions on $p$-adic curves of higher genus. 
These can be built using either sewing and cutting relations or the Schottky construction, which act on the tensor network in a natural way.
Since the tensor network reproduces the OPE, it is guaranteed to reproduce all correlation functions --- not only on the $p$-adic line, but on $p$-adic curves of arbitrary genus. 
Requiring the analogue of ``modular invariance'' however does not appear to lead to new constraints on the theory apart from an associative fusion algebra. 

A second application of our tensor networks is the study of RG flows. 
Since the tensor network describes the CFT path integral in the presence of general sources, it is straightforward to turn these sources on and study how they behave under a change of the UV cutoff. 
Of particular interest is the behavior of the fixed points of such RG flows.
Using our axioms, we proved that these fixed points always satisfy the axioms of $p$-adic CFT. 

One major departure from prior works on tensor network reconstruction is that our approach deals with the \emph{path integral} of the theory, rather than a wavefunction at some fixed time. 
The desirable properties of the p-adic tensor networks discussed \cite{Bhattacharyya:2017aly} are preserved, but within the new framework we can find the explicit component tensors needed to build any boundary $p$-adic CFT. 
This makes it the first precise and proven example of a \emph{tensor network/CFT
correspondence} in the sense of holographic tensor networks.
It may thereby be able to offer insights into the building blocks of holographically dual theories. 

There are some interesting features of our tensor network we only touched on briefly in the main text.
Bulk legs, which have played an important role in prior discussions of holographic tensor networks \cite{HaPPY,Hayden:2016cfa}, are simple to incorporate into our construction. 
In fact, they appeared briefly in our computation of the bulk 1-point function in the presence of sources in \secref{sec:correlators}. 
Bulk legs are added by replacing a bulk vertex tensor $T^{(q+1)}$ by $T^{(q+2)}$,  and any primary may be inserted on the new leg in the same manner as a boundary insertion.
These bulk legs are related to boundary legs by a version of the HKLL relation \cite{Hamilton:2005ju,Hamilton:2006az} along the lines described in \cite{Bhattacharyya:2017aly}. 

In the current tensor network construction, the structure constant of the $p$-adic CFT appears at every vertex. 
A very similar situation occurs in the study of Wilson line networks in AdS${}_{3}$/CFT${}_{2}$. 
There, Einstein gravity is reformulated as a gauge theory with gauge group $\SL(2,\mathbb{R})\times$SL$(2,\mathbb{R})$ \cite{Achucarro:1987vz,Witten:1988hc}, which is the conformal group, and the Wilson lines fall into representations of the gauge group. 
These representations can be labeled by the primaries of the CFT . When these Wilson lines meet at a vertex,  gauge invariance at the vertex requires the introduction of the structure constants of the CFT, which now plays the role of intertwiners  projecting the vertex to a singlet \cite{Castro:2018srf,Ammon:2013hba,Fitzpatrick:2016mtp,Besken:2016ooo}. The experience with AdS${}_{3}$/CFT${}_{2}$ therefore suggests that a similar construction is possible in the $p$-adic AdS/CFT. The gauge group is expected to be PGL$(2,\mathbb{K})$, while the representations can again be labeled by primaries of the dual CFT. In which case, the tensor network might emerge as the expectation value of a Wilson line tensor network. This analogy is indeed borne out in \cite{Hung:2018mcn}, offering an alternative interpretation of the tensor network constructed in the current paper. 

We end with the following observation. 
It is the special, and somewhat degenerate, properties of $p$-adic CFT that make its equivalence with a simple bulk tensor network possible. 
The primary source of that simplicity is that $p$-adic CFT has no descendant fields, even with respect to global conformal symmetry. 
The lack of descendants can be traced to the fact that all fields are 
valued in $\C$ but live on the $p$-adics, which forces all derivatives to vanish. 

A modification that may lead to a richer theory would be to require fields to be \emph{analytic functions}, something that is possible only if they are valued in a $p$-adic field. 
This assumption would not only introduce descendants with respect to global conformal symmetry, but may also distinguish between quasi-primary operators, which act as primaries under the global conformal algebra, and primary operators with respect to local analytic mappings, which are analogous to the Virasoro algebra. 
In particular, we would expect such a theory to have a stress tensor.
The presence of descendants would enforce stronger crossing relations on the CFT data, leading to a more rigid and richer theory that is more closely analogous to real CFTs. 
If such theories also have tensor network duals, it may lead to new insights into the structure of the AdS/CFT correspondence. 
These interesting questions are left for future exploration.

\section*{Acknowledgements} 
We thank Arpan Bhattacharyya and Long Cheng for initial collaboration, and Jie-qiang Wu for discussions. 
LYH thanks ITP-CAS and WL thanks Fudan University for hospitality during various stages of this project. 
We are grateful for support from the Thousand Young Talents Program. LYH is also supported by NSFC grant number 11875111. WL also thanks the support from Max-Planck Partergruppen fund and NSFC 11875064. 
The work of CMT was additionally supported by Fudan University and an Alexander von Humboldt Postdoctoral Research Fellowship.

\bibliographystyle{utphys}
\bibliography{pTN}

\end{document}